\documentclass[11pt,dvips]{article}
\usepackage{graphicx}
\usepackage{amssymb}
\usepackage{latexsym}
\usepackage{color}
\newcommand{\bra}{\begin{array}}
\newcommand{\era}{\end{array}}
\newcommand{\beq}{\begin{equation}}
\newcommand{\eeq}{\end{equation}}
\newcommand{\beqar}{\begin{eqnarray}}
\newcommand{\eeqar}{\end{eqnarray}}

\def\BC{\bb C}
\def\_\BC{\bbi C}

\def\( {\left(}
   \def\) {\right)}
\def\[ {\left[}
\def\] {\right]}
\def\no2 {{\textstyle{n\over 2}}}

\newcommand{\be}{\beta}

\newcommand{\al}{\alpha}

\newcommand{\da}{\dagger}

\begin{document}
\begin{titlepage}
\setcounter{page}{1}
\renewcommand{\thefootnote}{\fnsymbol{footnote}}

\begin{flushright}
%ucd-tpg 10.xx\\
%arXiv:yymm.xxxx
\end{flushright}

\vspace{5mm}
\begin{center}

{\Large \bf {Dirac Fermions in Inhomogeneous Magnetic
 Field}}
% and Klein Paradox }}
%Bloch Theory %in periodic structures
%Coupled to Rashba Interaction\\ and Gauge Field}

\vspace{5mm}

{\bf Ahmed Jellal$^{a,b,c}$\footnote{jellal@pks.mpg.de, ahjellal@kfu.edu.sa}}
and {\bf Abderrahim El Mouhafid$^{c}$\footnote{elmouhafid@gmail.com}}

\vspace{5mm}

{$^a$\em Physics Department, College of Science, King Faisal University,\\
PO Box 380, Alahsa 31982,
Saudi Arabia}

{$^b$\em Saudi Center for Theoretical Physics, Dhahran,
Saudi Arabia}

{$^{c}$\em Theoretical Physics Group,  %Department of Physics,
Faculty of Sciences, Choua\"ib Doukkali University},\\
{\em PO Box 20, 24000 El Jadida,
Morocco}\\

%\vspace{30mm}

\vspace{3cm}

\begin{abstract}

We study a confined system of  Dirac fermions in the presence
of inhomogeneous magnetic field. %s $B$ and $B'$.
Splitting the system into different regions, we determine
their corresponding energy spectrum solutions. We underline
their physical properties by considering the conservation energy
where some interesting relations are obtained. These are used to discuss
the reflexion and transmission coefficients for Dirac fermions and check
the probability condition for different cases. We generalize the obtained results
to a system with gap and make some analysis. After evaluating
the  current-carrying states, we analyze the Klein paradox and report interesting
discussions. % about three limiting cases.

% \pacs{02.30.Gp, 02,30.Tb, 02.30.Jr}

% \maketitle

%\newpage

\end{abstract}
\end{center}
\end{titlepage}

\newpage

%%%%%%%%%%%%%%%%%%%%%%%%%%%%%%%%%%%%%%%%%%%%%%%%%%
\section{Introduction}
%%%%%%%%%%%%%%%%%%%%%%%%%%%%%%%%%%%%%%%%%%%%%%%%%%%%%%

The Dirac formalism plies an important role not only
from mathematical point of view but physical one as well.
The recent observation of the anomalous quantum Hall effect
in graphene~\cite{EH0,EH01} renewed the interest to this formalism.
In fact, many
questions,  raised in graphene, found their solutions by adopting the Dirac formalism
as cornerstone.
Among them, we cite the confinement~\cite{EH02} that is much needed to describe
the transport properties in graphene. Subject that attracted much attention
where  interesting developments appeared by dealing
with different issues, for instance we refer to~\cite{peres,jellal1}.

%In this paper, we pose (and affirmatively answer)
On the other hand,
%the question whether
the quantum wires (electron waveguides) with quantized
conductance can be formed in graphene~\cite{mar}. Such
electron waveguides are indispensable parts of any conceivable
all-graphene device. In lithographically formed
graphene 'ribbons', the electronic bandstructure is theoretically
expected to very sensitively depend on the width
and on details of the boundary~\cite{3}. On top of that, disorder
and structural inhomogeneity are substantial in real
graphene~\cite{4}. For narrow graphene ribbons or electrostatically
formed graphene wires~\cite{5}, conventional conductance
quantization thus seems unlikely~\cite{6}. This expectation
is in accordance with recent experiments~\cite{7}.

%Contrary to such pessimism,
%by designing a suitable inhomogeneous magnetic field, it is demonstrated
%that a
%magnetic waveguide can be built that indeed allows for
%the perfectly quantized two-terminal conductance $4e^2/h$
%(including spin and valley degeneracy) even when disorder
%is present. The disorder insensitivity is based
%on a spatial separation of the left- and right-moving
%'snake' states found.
% under the model geometry shown
%in Fig. 1(a).
%This is reminiscent of the edge states encountered
%in the integer quantum Hall regime~\cite{8}, but
%in~\cite{mar} refers to a completely different microscopic picture.
%Such double-snake states develop in the regime B > 0 but
%B0 < 0, while an individual snake state is uni-directional
%and already found in the setup of Fig. 1(b).

Magnetic
barrier technology is well developed~\cite{9,10,11} and its
application to graphene samples appears to pose no fundamental
problems~\cite{mar}. In fact, snake states are experimentally
studied in other materials~\cite{9,13}, mainly motivated
by the quest for electrical rectification. On the
theory side, the confined Schr\"odinger fermions
in the magnetic field (with $B' = 0$) is discussed~\cite{14}
as well as the asymmetric cases~\cite{15}. For the Dirac-Weyl
quasiparticles encountered in graphene, however, such
calculations are not reported. The inhomogeneous magnetic
field case in graphene is analyzed in~\cite{16}.
%, and
%we employ that framework in our proposal of magnetic
%waveguides in graphene.
%On the other hand,
Theoretically, the electron waveguides,
in graphene created by suitable inhomogeneous magnetic fields, is considered~\cite{mar}.
The properties of uni-directional snake states are discussed. For a certain magnetic
field profile, two spatially separated counter-propagating snake states are formed, leading to conductance
quantization insensitive to backscattering by impurities or irregularities of the magnetic
field.

Subsequently, The tunneling effect of two-dimensional Dirac fermions in a constant magnetic field
is studied~\cite{cho}. This is done by using the continuity equation at fixed points to determine the
corresponding reflexion and transmission coefficients. For this, a system made of graphene,
as superposition of two different regions where the second is characterized by an energy gap t',
is considered. In fact,
concrete systems are treated to practically give two illustrations: barrier and diode. For each case,
the transmission in terms of the ratio of the energy conservation and t' are discussed.
Moreover, the resonant tunneling by introducing a scalar Lorentz potential is analyzed where
it is shown that a total transmission is possible.

Motivated by the above progress and in particular~\cite{mar,cho}, we deal with
other features of the system considered in~\cite{mar}. Such
system is composed of different regions submitted to two magnetic fields
and confined to a constant potential. This allows us to  treat each
region separately by determining the corresponding energy spectrum solutions. We underline some
physical properties of their spectrum by taking into account of
the energy conservation where interesting relations are obtained. Using the continuity
at different points, we explicitly evaluate the reflexion and transmission coefficients.
Combining all, we chow that the probability condition is well verified.
As second task, we consider the present system with energy gap $t'$ and do the same job to derive
its eigenspinors  as well as eigenvalues. It is shown that %As consequence, we conclude that
even the reflexion and transmission coefficients take new forms in terms of gap but the probability condition
still verified.  Interesting limits are discussed, which concern total reflexion
and transmission of the system with gap.

%In the last task
Finally, we  treat the Klein paradox by using  the  current-carrying states
 where different limits and discussions are presented. More precisely, we evaluate
 the currents for each region and use their relations to the reflexion and transmission
 coefficients to
  check the probabilities. Subsequently, three different cases are considered, which correspond to week, intermediate
  and strong potentials. We notice that two last cases are shown
   negative transmissions. However, by combining all coefficients
  we end up with a sum equal unity.

The present paper is organized as follows. In section 2, we
consider a confined Dirac fermion in inhomogeneous magnetic field (\ref{field}).
After getting the eigenvalues and eigenfunctions,
we analyze the energy conservation that allows us to derive interesting
relations between involved quantum numbers and parameters. In section 3,
we study scattering between two regions to determine the reflexion and
transmission coefficients, which will be used to discuss the probability
conditions of the present system. We do the same job in section 4 but by considering three
regions where the first is equivalent to
the third. The continuity at each point leads to
express the coefficients entering in the game
in terms of different parameters. In section 5, we introduce
a gap like a mass term and analyze the tunneling effect of
such case. We study the Klein paradox in section 6 by
involving the currents corresponding to different
regions and consider three cases. Finally, we close
by concluding our work.

%%%%%%%%%%%%%%%%%%%%%%%%%%%%%%%%%%%%%%%%%%%%%%%%%%%%%%%
\section{Dirac fermions in inhomogeneous magnetic field}
%%%%%%%%%%%%%%%%%%%%%%%%%%%%%%%%%%%%%%%%%%%%%%%%%%%%%%%

We consider  a system of  massless Dirac
fermions through a strip of graphene
characterized by the length $d$ and
width $W$ in the presence of inhomogeneous magnetic field. More precisely, we
introduce two magnetic fields $B$ and $B'$,
such as
%investigate the basic features
%of such system by considering a configuration of magnetic field,
%such as
\begin{eqnarray}\label{field}
 B(x)=\left\{%
\begin{array}{ll}
    B, & \qquad x<-d \\
    B', & \qquad |x|<d \\
    B, & \qquad x>d. \\
\end{array}%
\right.
\end{eqnarray}
According to the  configuration (\ref{field}), we decompose the present system into three regions.
Schematically, we end up with Figure 1

%%%%%%%%%%%%%%%%%%%%%figure%%%%%%%%%%%%%%%%%%%%%%%%%%%%%%
\begin{center}
% \begin{figure}
 \includegraphics[width=2.5in]{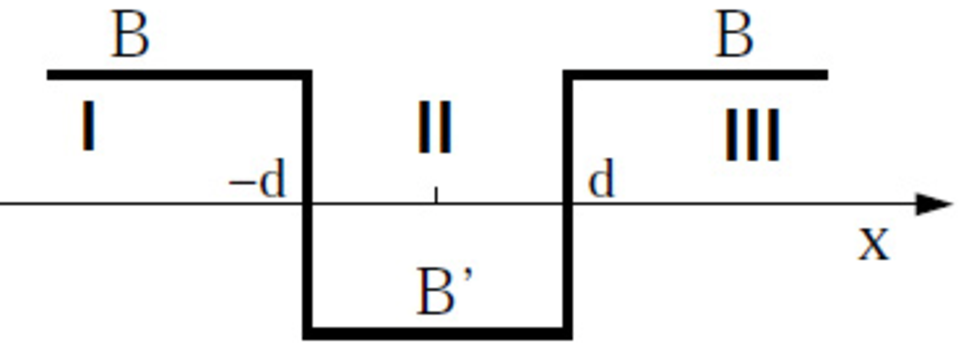}\\
 {\sf Figure 1}: {Magnetic field profile.}
%\end{figure}
\end{center}
%%%%%%%%%%%%%%%%%%%%%%%%%%%%%%%%%%%%%%%%%%%%%%%%%%%%%%%%%

\noindent Clearly, regions {\sf I} and {\sf III} are similar but different with respect to region {\sf II}.
Note that, the  system characterized by Figure 1 has been analyzed in~\cite{mar} for  possible quantum wires
in graphene. However,
in the present work we study other features of such system
to deal with different issues, which concern tunneling effect and
Klein paradox.

%Note that, this syetem has been studied in~\cite{mar}

%%%%%%%%%%%%%%%%%%%%%%%%%%%%%%%%%%%%%%%%%%%%%%%%%%%%%%%
\subsection{Dirac Hamiltonian}
%%%%%%%%%%%%%%%%%%%%%%%%%%%%%%%%%%%%%%%%%%%%%%%%%%%%%%%

Before writing down the appropriate Hamiltonian of the system (Figure 1), let us derive the corresponding
gauge field to the configuration (\ref{field}).
%Starting from the above configuration of magnetic field, we can
%deduce the corresponding gauge field.
Indeed, using the continuity
of the potential to obtain
\begin{eqnarray}
 A_{\sf j}(x)=\left\{%
\begin{array}{ll}
    A_{\sf I}(x)=Bx+(B-B')d, & \qquad x<-d\\
    A_{\sf II}(x)=B'x, & \qquad |x|<d \\
    A_{\sf III}(x)=Bx-(B-B')d, & \qquad x>d. \\
\end{array}%
\right.
\end{eqnarray}
where  ${\sf j}$ is labeling  regions {\sf I}, {\sf II} and {\sf III} . It clear that, for $B=B'$ we end up with
one potential and therefore three regions become similar to each others.

In the systems made of graphene, the
two Fermi points, each with a two-fold band degeneracy, can be described by a low-energy continuum
approximation with a four-component envelope wavefunction whose components are labeled by a
Fermi-point pseudospin $=\pm 1$ and a sublattice forming an honeycomb. Specifically, the Hamiltonian
for one-pseudospin component for the present system can be written as
%In each region, one massless  Dirac fermion can be described
%by the Hamiltonian
%like Hamiltonian describing three regions
%can be written in compact form as
\begin{eqnarray}\label{ak}
 H_{\sf j}=\upsilon_{F}\vec{\sigma}\cdot \vec{\pi}_{\sf j}+ V_{\sf j}(x)
\end{eqnarray}
where the components of the conjugate momentum
$\vec{\pi}_{\sf j}=\vec{p}+\frac{e}{c}\vec{A}_{\sf j}$  are given by
\begin{eqnarray}
\pi_{x,{\sf j}}=p_{x},
\qquad \pi_{y,{\sf j}}=p_{y}+\frac{e}{c}A_{\sf j}(x)
\end{eqnarray}
%and $p_{i}$, $i=x, y$, is the $i$ component of the momentum
%operator in the position basis $\textbf{p}=-i\hbar\nabla$.
and $V_{\sf j}(x)$ %, appearing in (\ref{ak}),
is the potential barrier that has a rectangular shape, which
is infinite along the $y$-axis and has the form
\begin{eqnarray}
 V_{\sf j}(x)=\left\{%
\begin{array}{ll}
    V_{0}, & \qquad -d<x<d \\
    0, & \qquad\hbox{otherwise} \\
\end{array}%
\right.
\end{eqnarray}
where $V_{0}>0$. Injecting all in (\ref{ak}) to get
\begin{eqnarray}\label{ham6}
H_{\sf j}=\upsilon_{F}\left(%
\begin{array}{cc}
  V_{\sf j}(x)/\upsilon_{F} & p_{x}-ip_{y}-i\frac{e}{c}A_{\sf j}(x) \\
  p_{x}+ip_{y}+i\frac{e}{c}A_{\sf j}(x) & V_{\sf j}(x)/\upsilon_{F} \\
\end{array}%
\right).
\end{eqnarray}

At this stage, it is convenient to
introduce the annihilation and creation operators. They can be defined
as
\begin{eqnarray}
a_{\sf j}=ip_{x}+p_{y}+\frac{e}{c}A_{\sf j}(x), \qquad
a_{\sf j}^{\dagger}=-ip_{x}+p_{y}+\frac{e}{c}A_{\sf j}(x)
\end{eqnarray}
which obey the canonical commutation relations
%\begin{eqnarray}
 %[a_{\sf j},a_{\sf j}^{+}]=-\frac{2ie}{c}\left[A_{\sf j}(x),p_{x}\right].
%\end{eqnarray}
%Explicitely, they are given by
\begin{eqnarray}
 \left[a_{\sf j},a_{\sf j}^{\dagger}\right]=\left\{%
\begin{array}{ll}
    \frac{2\hbar^{2}}{l_{B}^{2}}, & \qquad j={\sf I} \\
    \frac{2\hbar^{2}}{l_{|B'|}^{2}}, & \qquad j={\sf II} \\
    \frac{2\hbar^{2}}{l_{B}^{2}}, & \qquad j={\sf III} \\
\end{array}%
\right.
\end{eqnarray}
where the magnetic lengths $l_{B}=\sqrt{\frac{\hbar c}{eB}}$ and
$l_{|B'|}=\sqrt{\frac{\hbar c}{e|B'|}}$ are corresponding to the magnetic fields $B$ and
$B'$, respectively. The Hamiltonian (\ref{ham6}) can be written in terms of $a_{\sf j}$ and $a_{\sf j}^{\da}$  as
\begin{eqnarray}\label{ham9}
H_{\sf j}=iv_{F}\left(%
\begin{array}{cc}
  V_{\sf j}(x)/i\upsilon_{F} & -a_{\sf j} \\
  a_{\sf j}^{\dagger} & V_{\sf j}(x)/i\upsilon_{F} \\
\end{array}%
\right)
\end{eqnarray}
which is encoding all regions. This will be used to  study
each region separately  and derive the corresponding energy spectrum solutions.

%%%%%%%%%%%%%%%%%%%%%%%%%%%%%%%%%%%%%%%%%%%%%%%%%%%%%%%
\subsection{Energy spectrum solutions}
%%%%%%%%%%%%%%%%%%%%%%%%%%%%%%%%%%%%%%%%%%%%%%%%%%%%%%%

We determine the eigenvalues and eigenspinors of the Hamiltonian
$H_{\sf j}$. Indeed, the
 the Dirac Hamiltonian describing region {\sf I} is obtained from
 (\ref{ham9}) as
\begin{eqnarray}
H_{\sf I} %=H_{K_{y}}^{B}
=iv_{F}\left(%
\begin{array}{cc}
 0 & -a_{\sf I} \\
  a_{\sf I}^{\dagger} & 0 \\
\end{array}%
\right).
\end{eqnarray}
The operators $a_{\sf I}$ and $a_{\sf I}^{\dagger}$ can be rescaled to
define others, such as
%We consider new operators in terms of $a_{1}$ and $a_{1}^{+}$.
%They are
\begin{eqnarray}
b_{\sf I}=\frac{l_{B}}{\sqrt{2}\hbar}a_{\sf I}, \qquad
b_{\sf I}^{\dagger}=\frac{l_{B}}{\sqrt{2}\hbar}a_{\sf I}^{\dagger}
\end{eqnarray}
which verify
\begin{eqnarray}
 [b_{\sf I},b_{\sf I}^{\dagger}]=\mathbb{I}.
\end{eqnarray}
Using these to write $H_{\sf I}$ as
\begin{eqnarray}\label{1}
H_{\sf I}=i\hbar\omega_{c}\left(%
\begin{array}{cc}
  0 & -b_{\sf I} \\
  b_{\sf I}^{\dagger} & 0 \\
\end{array}%
\right)
\end{eqnarray}
where we have set $\omega_{c}=\sqrt{2}\frac{v_{F}}{l_{B}}$ as a cyclotron
frequency.

To get the energy spectrum solutions of (\ref{1}), we solve
the eigenvalue equation for a given spinor
$\phi_{\sf I}=\left(\begin{array}{c}
  \varphi_{1} \\
  \varphi_{2} \\
\end{array}\right)$ of $H_{\sf I}$. This is
\begin{eqnarray}
H_{I}\left(\begin{array}{c}
  \varphi_{1} \\
  \varphi_{2} \\
\end{array}\right)=E_{\sf I}\left(\begin{array}{c}
  \varphi_{1} \\
  \varphi_{2} \\
\end{array}\right)
\end{eqnarray}
which is equivalent to
\begin{eqnarray}
i\hbar\omega_{c}\left(%
\begin{array}{cc}
  0 & -b_{\sf I} \\
  b_{\sf I}^{\dagger} & 0 \\
\end{array}%
\right)\left(\begin{array}{c}
  \varphi_{1} \\
  \varphi_{2} \\
\end{array}\right)=E_{\sf I}\left(\begin{array}{c}
  \varphi_{1} \\
  \varphi_{2} \\
\end{array}\right)
\end{eqnarray}
%where the components of the Dirac spinor $\varphi_{1}$ and
%$\varphi_{2}$ are given by
and leads to two relations between spinor components
\begin{eqnarray}
-i\hbar\omega_{c}b_{\sf I}\varphi_{2}=E_{\sf I}\varphi_{1}\label{2}\\
i\hbar\omega_{c}b_{\sf I}^{\dagger}\varphi_{1}=E_{\sf I}\varphi_{2}.\label{3}
\end{eqnarray}
Now inserting (\ref{2}) into (\ref{3}) to obtain a differential equation of
second order for $\varphi_{2}$
\begin{eqnarray}\label{4}
\hbar^{2}\omega_{c}^{2}b_{\sf I}^{\dagger}b_{\sf I}\varphi_{2}=E_{\sf I}^{2}\varphi_{2}.
\end{eqnarray}
It is clear that $\varphi_{2}$ is an eigenstate
of the number operator $\hat{n}=b_{\sf I}^{\dagger}b_{\sf I}$ and therefore we identify
$\varphi_{2}$ to the eigenstates of the harmonic oscillator  $\mid n\rangle$, namely
\begin{equation}\label{eigs1}
\varphi_{2}\sim\mid n\rangle
\end{equation}
 and its eigenvalues read as
\begin{eqnarray}\label{5}
E_{{\sf I},n}=s \hbar \omega_{c}\sqrt{|n|} %=s\frac{3ta}{2}\sqrt{eB|n|}
\end{eqnarray}
where $n$ is obviously the eigenvalues of $\hat{n}$, with $n= {0,
\pm1, \pm2, \cdots}$, and $s={\sf sgn}(n)$. Note that, $n=0$
corresponds to the lowest Landau level, i.e. zero mode energy.

Using (\ref{2}), (\ref{eigs1}) and (\ref{5}) to get the first component as
\begin{eqnarray}
\varphi_{1}=\frac{-i\hbar\omega_{c}}{E_{\sf I}}b_{\sf I}\mid n\rangle=-is\mid n-1\rangle
\end{eqnarray}
which gives the eigenspinors
\begin{eqnarray}
\phi_{{\sf I},n}\sim \left(%
\begin{array}{c}
  -s i \mid n-1\rangle \\
 \mid n\rangle \\
\end{array}%
\right).
\end{eqnarray}
In terms of the parabolic cylinder functions $D_{n}(x+x_{01})$~\cite{gra},
the eigenspinors in the plane $(x,y)$ are
\begin{eqnarray}\label{6}
\phi_{{\sf I},n,k_{y}}(x,y)=\frac{1}{\sqrt{2}}\left(%
\begin{array}{c}
  -si D_{|n|-1}(x+x_{01}) \\
 D_{|n|}(x+x_{01}) \\
\end{array}%
\right)\ e^{ik_{y}y}
\end{eqnarray}
where  $x_{01}=k_{y}l_{B}^{2}+\left(1-\frac{|B'|}{B}\right)d$ and
$D_{n}(x)$ are related to Hermite polynomials via
\begin{equation}
D_{n}(x)=\left(l_{B}\sqrt{\pi}n!2^{n}\right)^{-\frac{1}{2}}\exp\left(-\frac{x^{2}}{2}\right)H_{n}(x).
\end{equation}

As far as region {\sf II} is concerned,
%Similarly, the eigenfunctions for second region can be expressed
%in terms of $x_{02}=q_{y}l_{|B'|}^{2}$. %Otherwise, by replacing the quantities
we use the mapping $n\longrightarrow m$
and $s \longrightarrow s'$ in (\ref{6}) to obtain the eigenspinors
$\phi_{{\sf II},m,k_{y}}(x,y)$ in terms of $x_{02}=q_{y}l_{|B'|}^{2}$
and
the corresponding eigenvalues
\begin{eqnarray}\label{reg2sp}
E_{{\sf II},m}=s'\hbar\omega_{c}'\sqrt{|m|}+V_{0}
\end{eqnarray}
where $s'={\sf sgn}(m)$ and $\omega_{c}'=\sqrt{2}\frac{v_{F}}{l_{|B'|}}$ is
the cyclotron frequency associated to the magnetic field $B'$.

Finally, for  region {\sf III} the eigenvalues and the
eigenspinors are similar to those of region {\sf I} except that the
correspondence
\begin{equation}
x_{01} \longrightarrow
x_{03}=k_{y}l_{B}^{2}-\left(1-\frac{|B'|}{B}\right)d
\end{equation}
must be taken into account in (\ref{6}). Note that,
for $B'=-B$ (with $B'<0$) the eigenspinors for three regions can
be expressed with the same position $x_{0}=k_{y}l_{B}^{2}$.

%%%%%%%%%%%%%%%%%%%%%%%%%%%%%%%%%%%%%%%%%%%%%%%%%%%
\subsection{Illustrations}
%%%%%%%%%%%%%%%%%%%%%%%%%%%%%%%%%%%%%%%%%%%%%%%%%%%%

To give some illustrations, we focus on the eigenfunctions of
 four lowest states in region $\sf II$, which are summarized
in Table 1. Different plots are given in Figure 2 those show that
 the $m^{th}$ eigenfunction has \emph{m}-nodes, namely
there are $m$-values of $x$ for which $\phi_{{\sf II},m}(x)=0$.
\begin{center}
\begin{tabular}{|c|l|l|}
   \hline
   ~Number~ &  ~Energy eigenvalue~ & ~~~~~~~~~~~~~~~ ~~~~~~~~ Energy eigenfunction \\
   \hline %\hline
   $m=0$ & $E_{{\sf II},0}=V_{0}$ & $\phi_{{\sf II},0}(x)=\left(\frac{1}{l_{|B'|}\sqrt{\pi}}\right)^{\frac{1}{2}}e^{-\left(x+x_{02}\right)^{2}/2l_{|B'|}^{2}}$ \\
   \hline
   $m=1$ & $E_{{\sf II},1}=\hbar\omega_{c}'+V_{0}$ & $\phi_{{\sf II},1}(x)=\left(\frac{1}{2l_{|B'|}\sqrt{\pi}}\right)^{\frac{1}{2}}
   2\left(\frac{x+x_{02}}{l_{|B'|}}\right)e^{-\left(x+x_{02}\right)^{2}/2l_{|B'|}^{2}}$ \\
   \hline
   $m=2$ & $E_{{\sf II},2}=\sqrt{2}\hbar\omega_{c}'+ V_{0}$ & $\phi_{{\sf II},2}(x)=\left(\frac{1}{8l_{|B'|}\sqrt{\pi}}\right)^{\frac{1}{2}}
   \left[4\left(\frac{x+x_{02}}{l_{|B'|}}\right)^{2}-2\right]e^{-\left(x+x_{02}\right)^{2}/2l_{|B'|}^{2}}$ \\
   \hline
   $m=3$ & $E_{{\sf II},3}=\sqrt{3}\hbar\omega_{c}'+ V_{0}$ & $\phi_{{\sf II},3}(x)=\left(\frac{1}{48l_{|B'|}\sqrt{\pi}}\right)^{\frac{1}{2}}
   \left[8\left(\frac{x+x_{02}}{l_{|B'|}}\right)^{3}-12\left(\frac{x+x_{02}}{l_{|B'|}}\right)\right]e^{-\left(x+x_{02}\right)^{2}/2l_{|B'|}^{2}}$ \\
   \hline
\end{tabular}
\end{center}
\begin{center}
{\sf Table 1}: Normalized eigenfunctions for four lowest states of a
one-dimensional potential energy field.
% where the parameter
%$l_{|B'|}=\sqrt{\frac{\hbar c}{e|B'|}}$
%determines the spatial extent of the eigenfunctions.
\end{center}

\begin{center}
\includegraphics[width=2.6in]%[width=-0.5cm,keepaspectratio]
{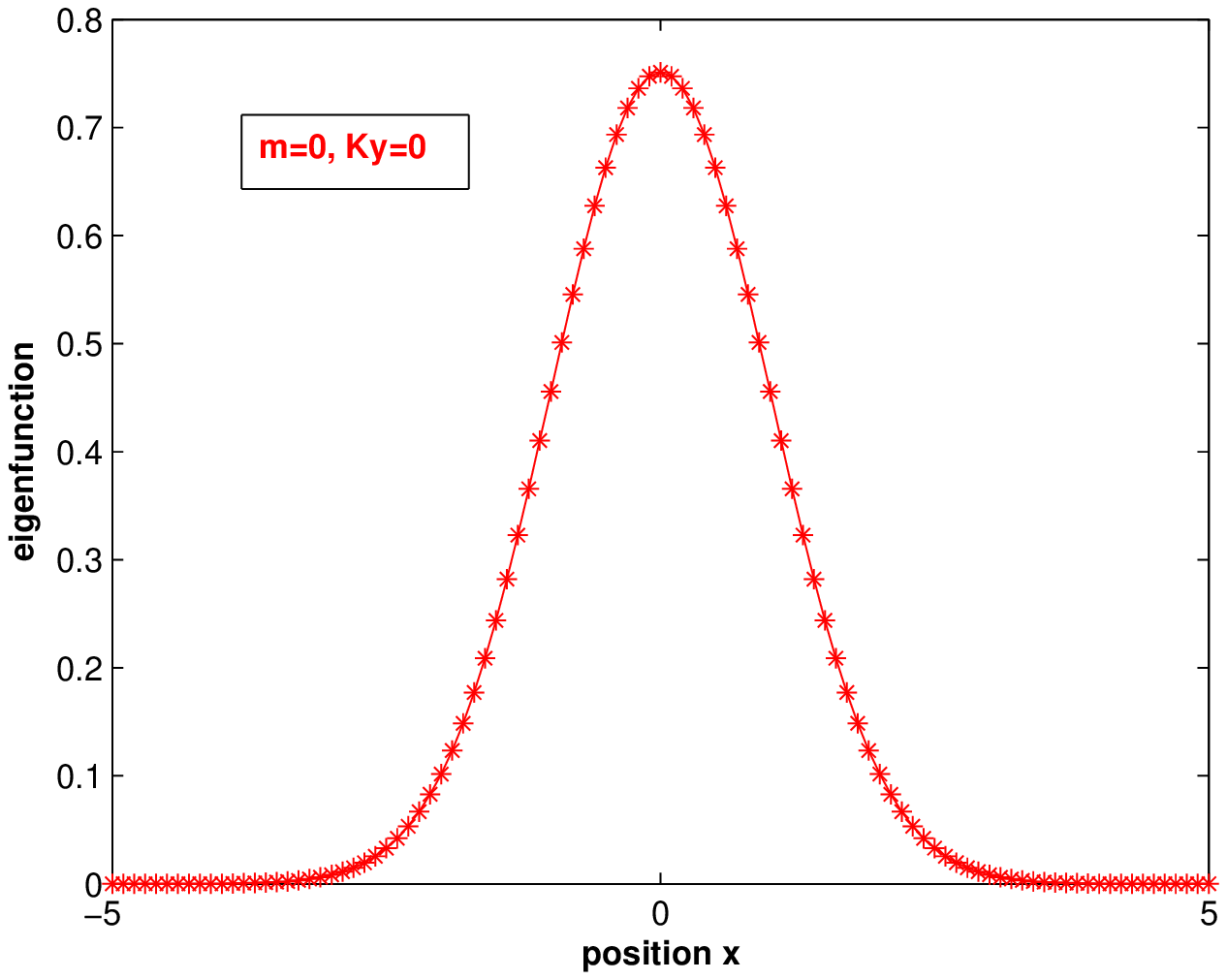}
%\end{figure}
 %\includegraphics[width=2.6in]{ihe9}
 \includegraphics[width=2.6in]%[width=-0.5cm,keepaspectratio]
{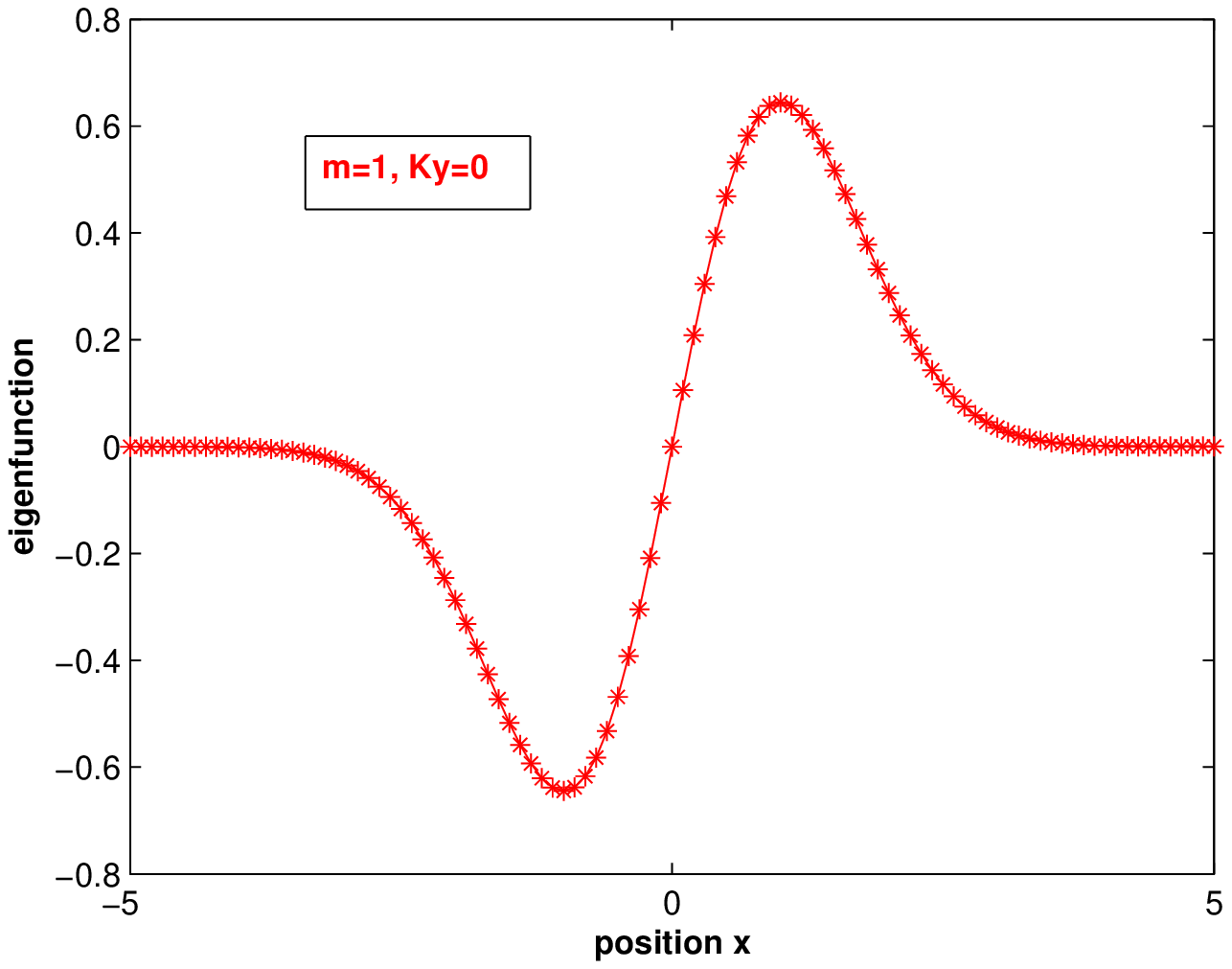}
%\end{figure}
 %\includegraphics[width=2.6in]{iheç9}
\end{center}
\begin{center}
\includegraphics[width=2.6in]%[width=-0.5cm,keepaspectratio]
{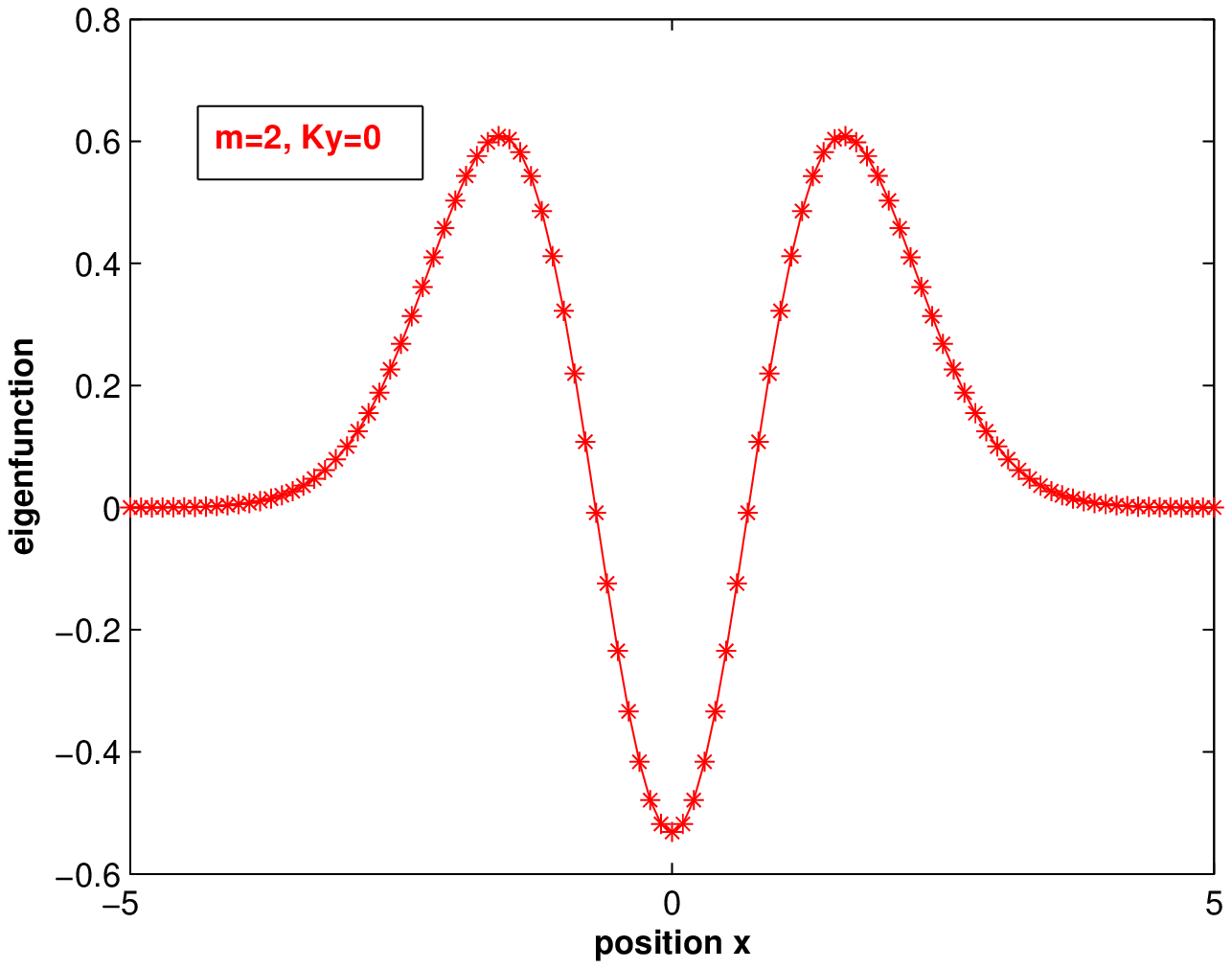}
%\end{figure}
 %\includegraphics[width=2.6in]{ihe9}
 \includegraphics[width=2.6in]%[width=-0.5cm,keepaspectratio]
{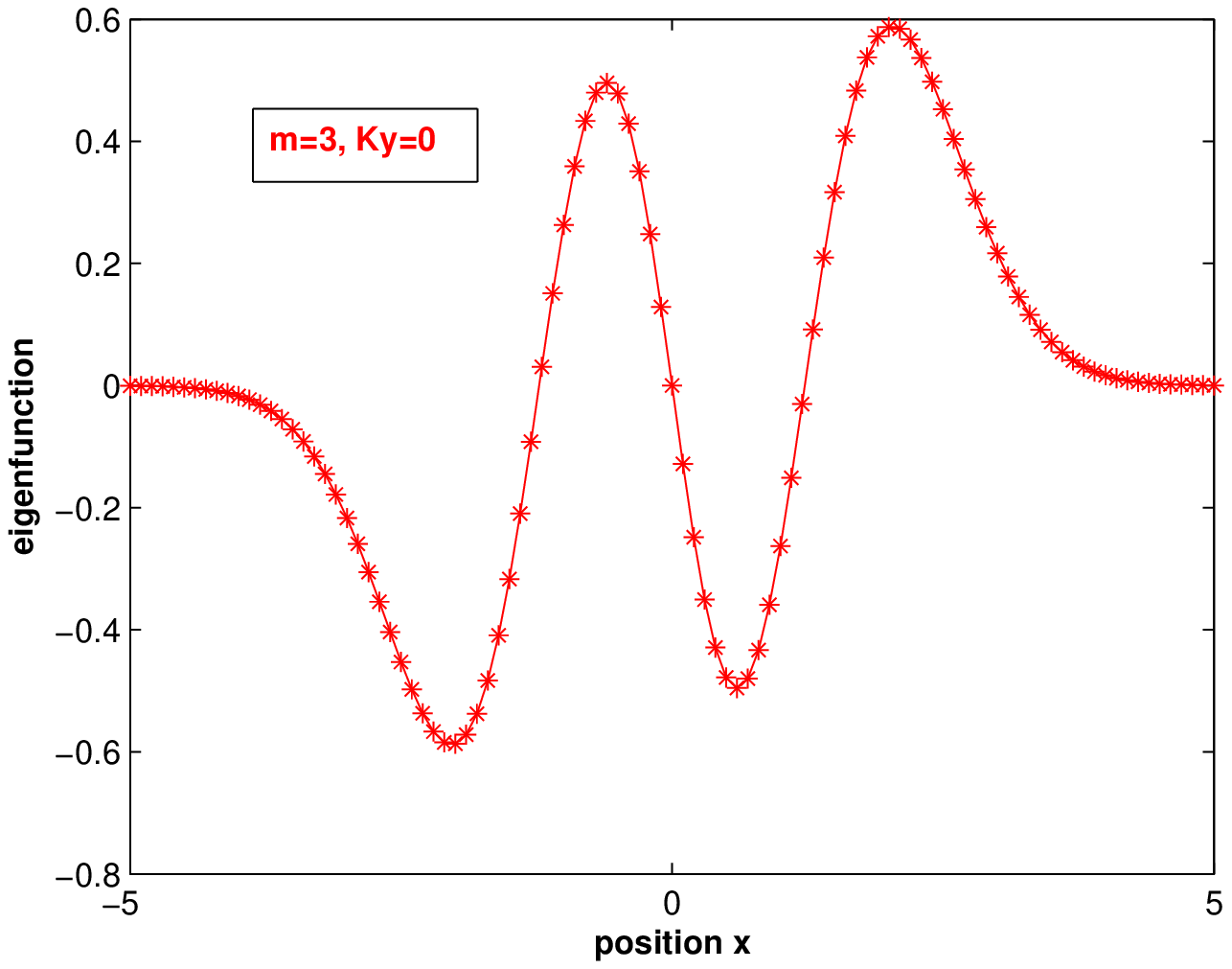}
%\end{figure}
 %\includegraphics[width=2.6in]{iheç9}
\end{center}
\begin{center}
{\sf Figure 2}: The spatial shapes of the eigenfunctions $\phi_{{\sf II},m}(x)$, with $m=0,1,2,3$
%the four lowest states of a one-dimensional potential energy field
%with length parameter
and $l_{|B'|}=\sqrt{\frac{\hbar c}{e|B'|}}=1$.
\end{center}

On the other hand,
the wavefunctions have observable properties. Indeed,
if the
position coordinate is changed from \emph{x} to \emph{-x}, the
eigenfunction has a definite symmetry
\begin{eqnarray}
\left\{%
\begin{array}{ll}
    \phi_{{\sf II},m}(-x)=+\phi_{{\sf II},m}(x), \qquad \hbox{if \emph{m} is even} \\
    \phi_{{\sf II},m}(-x)=-\phi_{{\sf II},m}(x), \qquad \hbox{if \emph{m} is odd} \\
\end{array}%
\right.
\end{eqnarray}
which is nothing but the parity symmetry. Furthermore,
%Oscillator states with even \emph{m} are said to have positive
%parity and oscillator states with odd \emph{m} are said to have
%negative parity.
the position probability density of fermion
is given by
\begin{eqnarray}
|\phi_{m}(x,y)|^{2}=|\phi_{m}(x)|^{2}.
\end{eqnarray}
By plotting this for some specific values of $m$, we deduce
an interesting conclusion. %In fact, by
%looking at
%As can be seen from
From Figure 3, we notice that the fermion can have any location
between $x=-\infty$ and $x=+\infty$, in marked contrast with a
classical fermion, which is confined to the region $-A<x<+A$
where $A$ is the amplitude of oscillation.
\begin{center}
\includegraphics[width=2.6in]%[width=-0.5cm,keepaspectratio]
{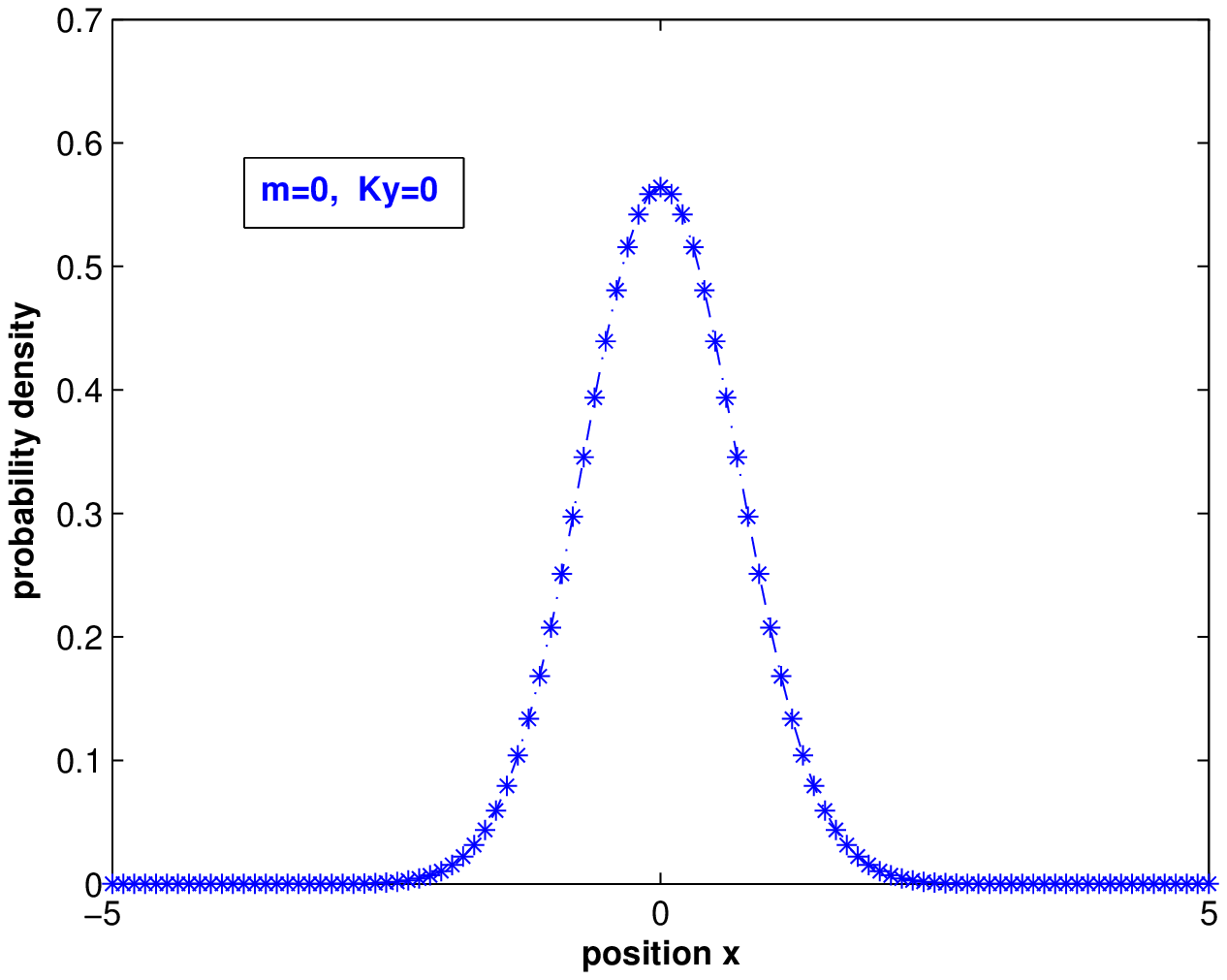}
%\end{figure}
 %\includegraphics[width=2.6in]{probkzero}
 \includegraphics[width=2.6in]%[width=-0.5cm,keepaspectratio]
{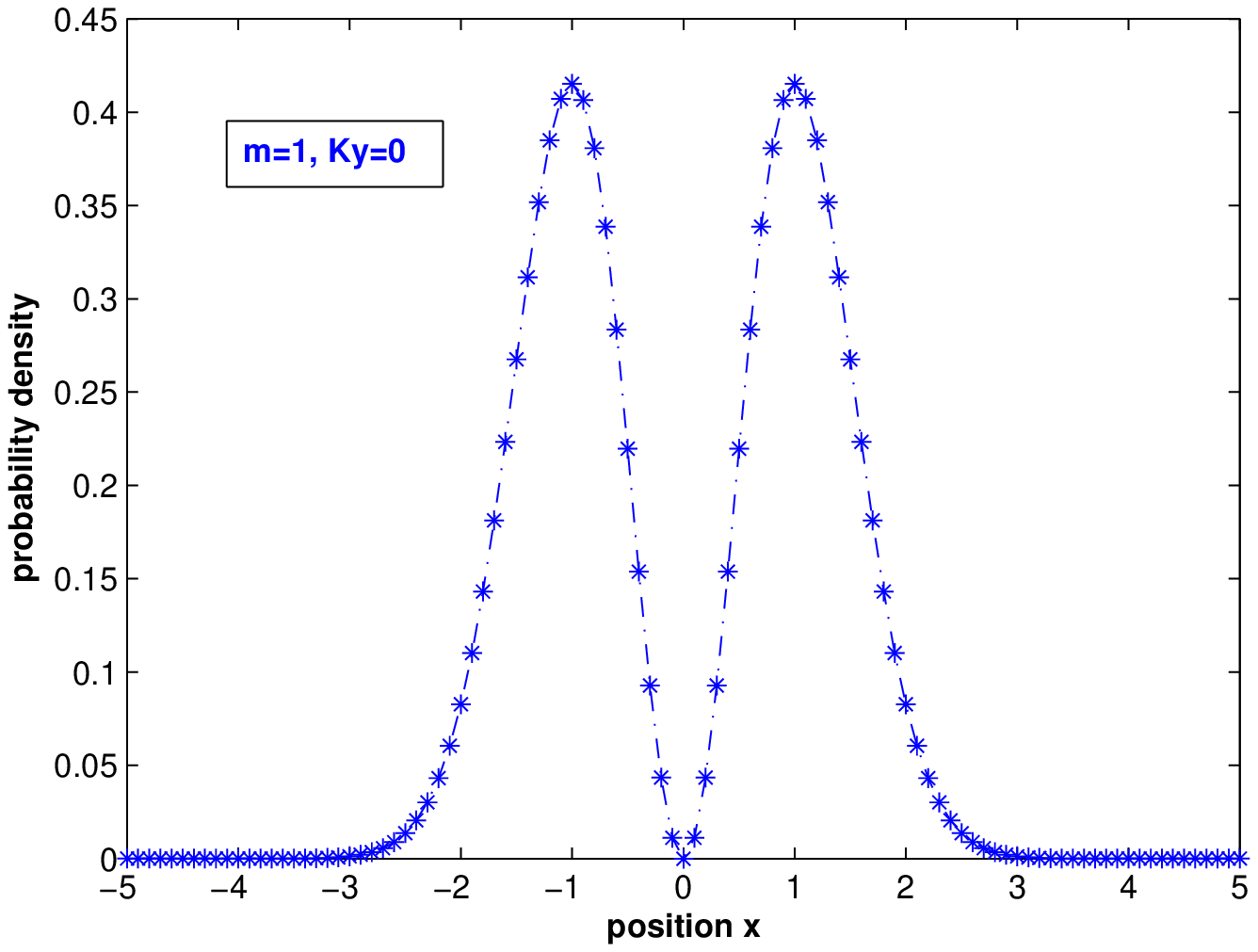}
%\end{figure}
 %\includegraphics[width=2.6in]{probkzerom1}
\end{center}
\begin{center}
\includegraphics[width=2.6in]%[width=-0.5cm,keepaspectratio]
{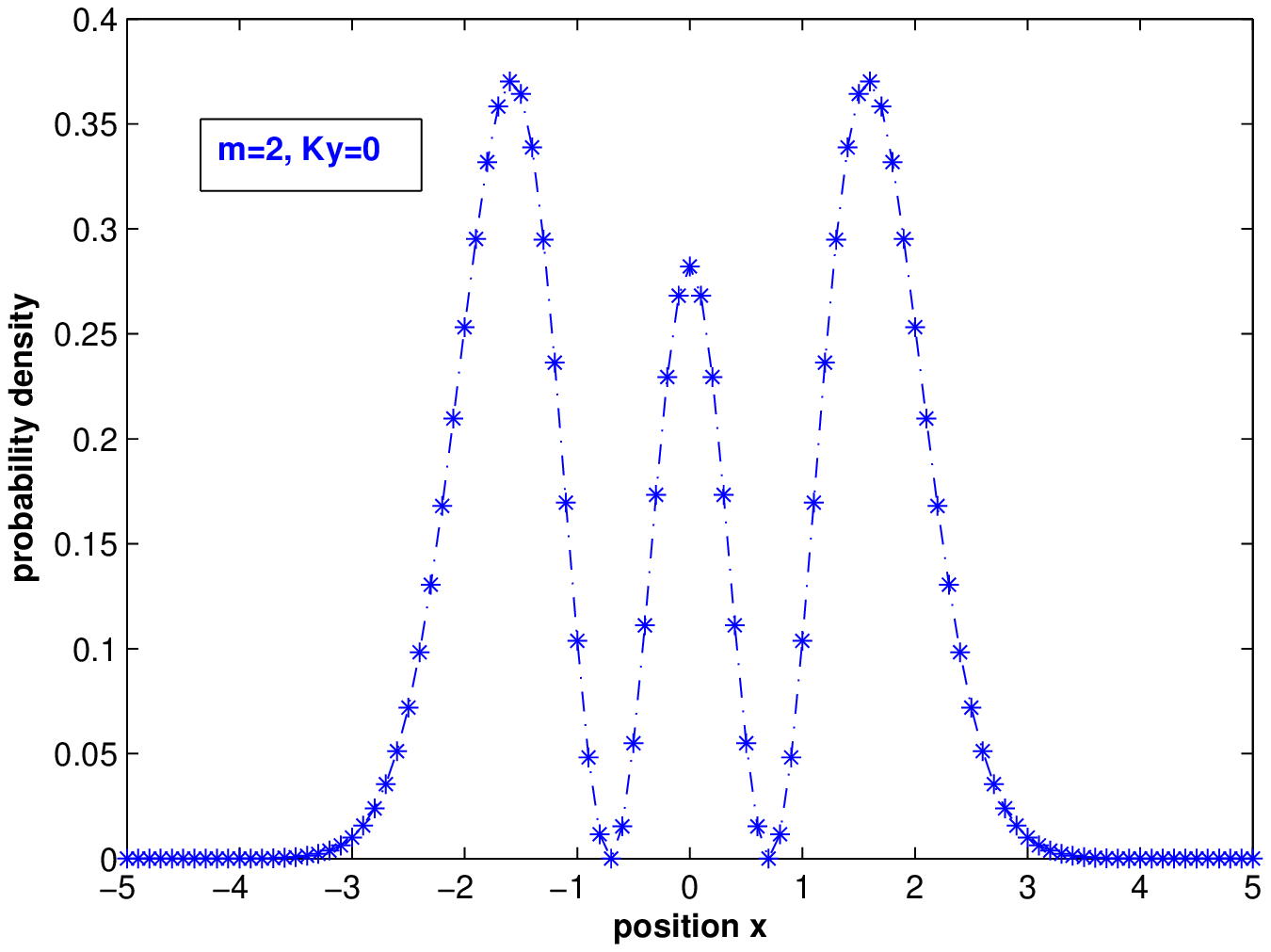}
%\end{figure}
 %\includegraphics[width=2.6in]{probkzerom2}
 \includegraphics[width=2.6in]%[width=-0.5cm,keepaspectratio]
{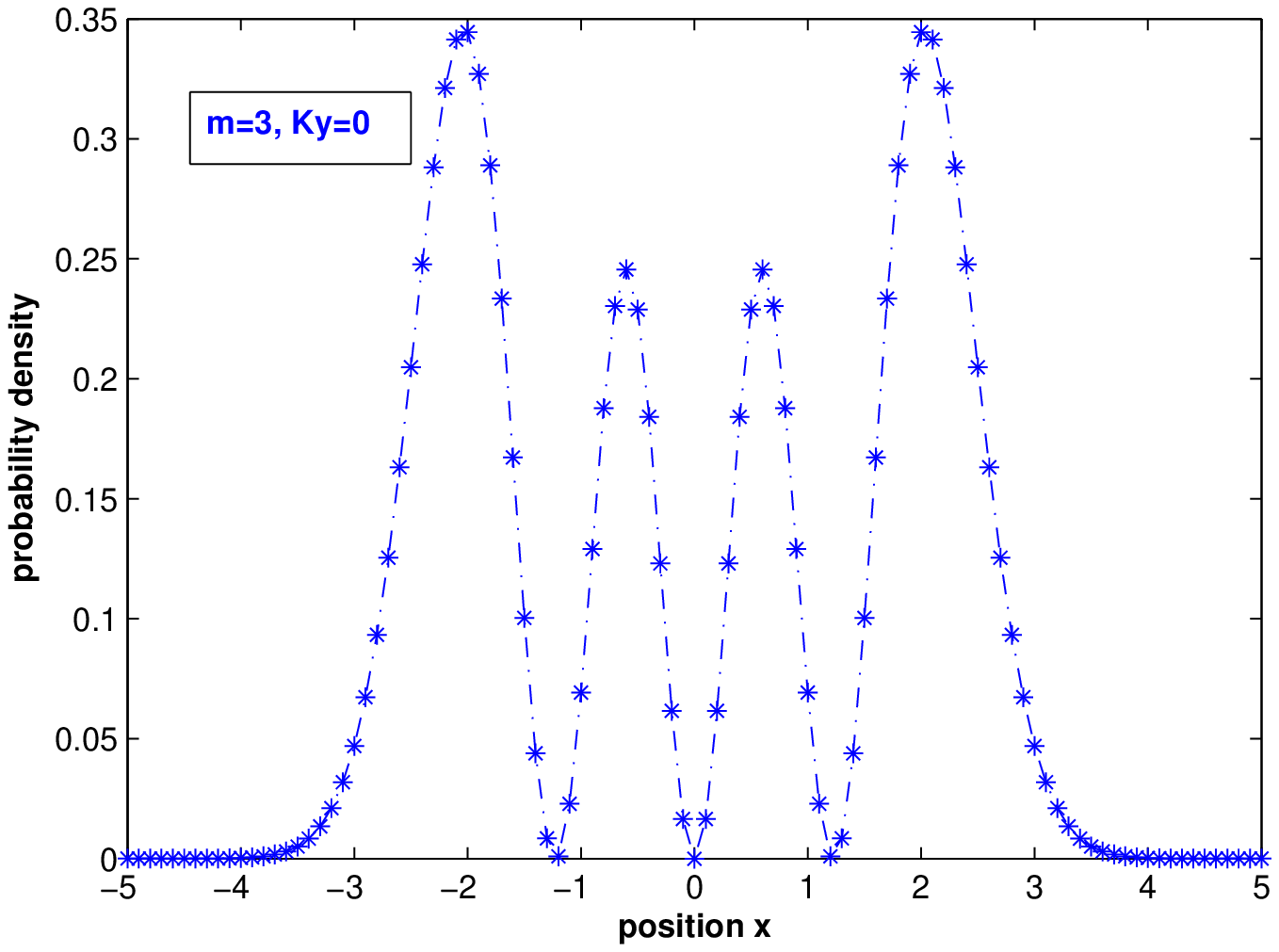}
%\end{figure}
 %\includegraphics[width=2.6in]{probkzerom3}
\end{center}
\begin{center}
{\sf Figure 3}: The position probability densities $|\phi_{{\sf II},m}(x)|^{2}$ corresponding to Table 1.
\end{center}

%%%%%%%%%%%%%%%%%%%%%%%%%%%%%%%%%%%%%%%%%%%%%%%%%%%%%%%%%
\subsection{Energy conservation}
%%%%%%%%%%%%%%%%%%%%%%%%%%%%%%%%%%%%%%%%%%%%%%%%%%%%%%%%

In the interface between  regions, there is
conservation of the tangent components of the wave vector, i.e.
$k_{y}=q_{y}$, and conservation of the energy. This is
\begin{eqnarray}
E_{\sf I}=E_{\sf II}=E
\end{eqnarray}
which leads to
the constraint
\begin{eqnarray}
\frac{|n|}{|m|}=\frac{|B'|}{B}\frac{E^{2}}{\left(E-V_{0}\right)^{2}}.
\end{eqnarray}
Since  $n$ and $m$ are integer values,
the r.h.s term must be a fractional number, which can be written as % Thus, one can write
%where the transverse quantization quantum numbers $n$ and $m$
%obeys the condition
\begin{equation}
|n|=K\ |m|, \qquad K\in {\mathbb{Q}}^{+}.
\end{equation}
This relation  is very important because without such set one can not talk
about tunneling effect in the present case. We will clarify this
statement  from next section and exactly when we begin by calculating
different quantities in order to check the probability condition.

%\textbf{Discussions}:
Now let us return to (\ref{5}) and (\ref{reg2sp}) to write the ratio as
\begin{eqnarray}\label{1encon}
\frac{E}{V_{0}}=\frac{\sqrt{|n|}}{\sqrt{|n|}-\frac{s'}{s}\sqrt{\frac{|B'|}{B}|m|}}.
\end{eqnarray}
Recall that
in the region ${\sf II}$ we have $V_{0}>0$, which implies that the
energy
$E$ can be either positive or negative. Therefore,
we should distinguish between two situations as listed
below\\
\begin{center}
\begin{tabular}{|r|l|}
\hline
$E>0 \ (s=s'=+1)$ & $E<0 \ (s=s'=-1)$ \\
\hline
$|n|>\frac{|B'|}{B}|m|$~~~~~~~~~~ & $|n|<\frac{|B'|}{B}|m|$ \\
\hline
\end{tabular}\\
\end{center}
\begin{center}
{\sf Table 2:} Positive and negative energies and their corresponding
quantum number configurations.
\end{center}
These energies can be plotted to explicitly illustrate
their behavior in terms of different quantities entering in the game, which are given in
Figure 4.

\begin{center}
\includegraphics[width=2.50in]%[width=-0.5cm,keepaspectratio]
{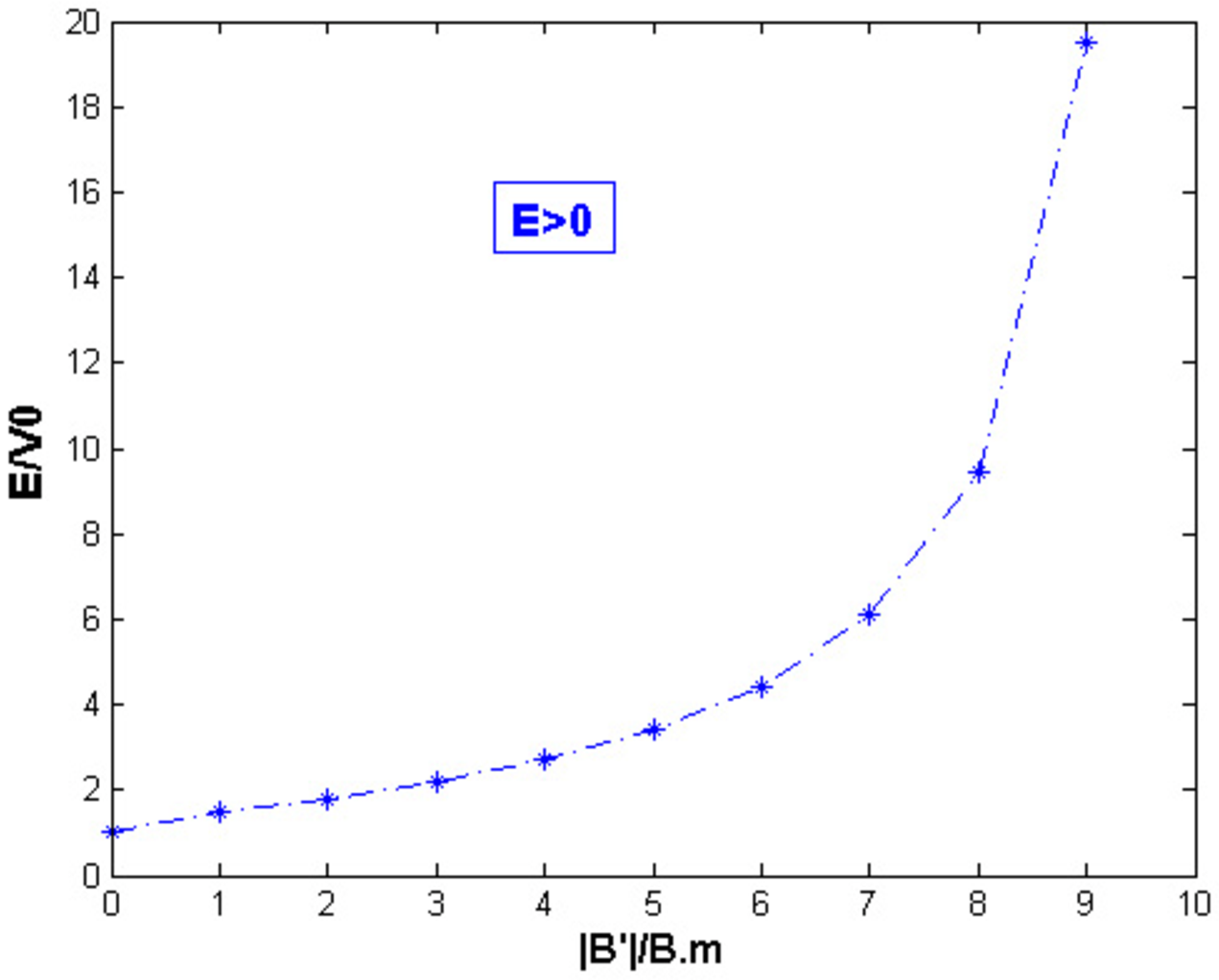}
%\end{figure}
 %\includegraphics[width=1.5in]{ihe9}
 \includegraphics[width=2.55in]%[width=-0.5cm,keepaspectratio]
{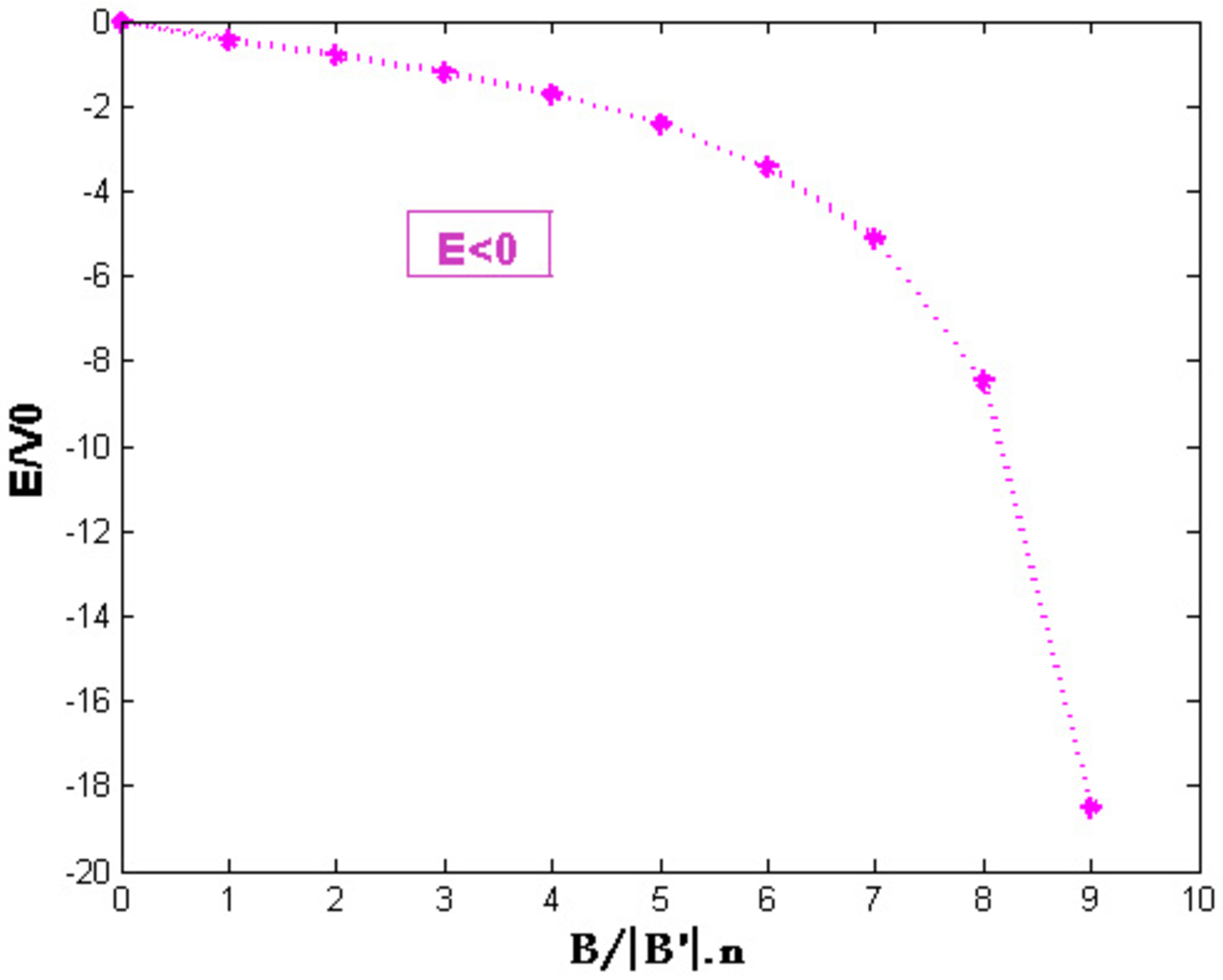}
%\end{figure}
 %\includegraphics[width=1.5in]{iheç9}
\end{center}
\begin{center}
{\sf Figure 4}: Variation of the ratio $\frac{E}{V_{0}}$
in terms of $\frac{|B'|}{B}|m|$ for $E>0$ and $\frac{B}{|B'|}|n|$ for $E<0$.
\end{center}

%%%%%%%%%%%%%%%%%%%%%%%%%%%%%%%%%%%%%%%%%%%%%%%%%%%%%%%
\section{Two regions in inhomogeneous magnetic field}
%%%%%%%%%%%%%%%%%%%%%%%%%%%%%%%%%%%%%%%%%%%%%%%%%%%%%%%

To treat a concrete example of the present system, we consider a
barrier  submitted to an inhomogeneous magnetic field. This barrier can be seen as
superposition of two regions separated by an interface localized
at a fixed point. We study the
tunneling effect by evaluating the reflexion and transmission
coefficients at interface, which in our case corresponds to the point zero.
These will be used to show that the probability condition is
exactly one and emphasis what makes difference with respect to without confinement
case~\cite{cho}. To do this task, we
follow the same lines as has been done~\cite{cho} and distinguish between propagation with positive and negative
incidences. In both cases we deal with propagation
from region {\sf I} to {\sf II}, {\sf II} to {\sf III} and vice verse.

%%%%%%%%%%%%%%%%%%%%%%%%%%%%%%%%%%%%%%%%%%%%%%%%%%%%%%%
\subsection{Propagation with positive incidence}
%%%%%%%%%%%%%%%%%%%%%%%%%%%%%%%%%%%%%%%%%%%%%%%%%%%%%%%

To proceed, let us first define
the eigenspinors for three regions in positive and negative
direction of the variable $x$. In region {\sf I}, we write
\begin{eqnarray}\label{7}
\phi_{{\sf I},+}=\frac{1}{\sqrt{2}}\left(%
\begin{array}{c}
  -si D_{|n|-1}(x+x_{01}) \\
 D_{|n|}(x+x_{01}) \\
\end{array}%
\right)e^{ik_{y}y}, \qquad
\phi_{{\sf I},-}=\frac{1}{\sqrt{2}}\left(%
\begin{array}{c}
  -si D_{|n|-1}(-x-x_{01}) \\
 D_{|n|}(-x-x_{01}) \\
\end{array}%
\right)e^{ik_{y}y}.
\end{eqnarray}
Similarly, in region {\sf II} we have
\begin{eqnarray}\label{8}
\phi_{{\sf II},+}=\frac{1}{\sqrt{2}}\left(%
\begin{array}{c}
  -s'i D_{|m|-1}(x+x_{02}) \\
 D_{|m|}(x+x_{02}) \\
\end{array}%
\right)e^{ik_{y}y}, \qquad
\phi_{{\sf II},-}=\frac{1}{\sqrt{2}}\left(%
\begin{array}{c}
  -s'i D_{|m|-1}(-x-x_{02}) \\
D_{|m|}(-x-x_{02}) \\
\end{array}%
\right)e^{ik_{y}y}
\end{eqnarray}
as well as in region {\sf III}
\begin{eqnarray}
\phi_{{\sf III},+}=\frac{1}{\sqrt{2}}\left(%
\begin{array}{c}
  -si D_{|n|-1}(x+x_{03}) \\
 D_{|n|}(x+x_{03}) \\
\end{array}%
\right)e^{ik_{y}y}, \qquad
\phi_{{\sf III},-}=\frac{1}{\sqrt{2}}\left(%
\begin{array}{c}
  -si D_{|n|-1}(-x-x_{03}) \\
 D_{|n|}(-x-x_{03}) \\
\end{array}%
\right)e^{ik_{y}y}
\end{eqnarray}
where $(\pm)$ refer to the positive and negative propagations, respectively.

%%%%%%%%%%%%%%%%%%%%%%%%%%%%%%%%%%%%%%%%%%%%%%%%%%%%%%%%%%%%%%%%%%%
%\subsubsection{Propagation from region {\sf I} to region {\sf II}}
%%%%%%%%%%%%%%%%%%%%%%%%%%%%%%%%%%%%%%%%%%%%%%%%%%%%%%%%%%%%%%%%%%%

We start by analyzing the case of {propagation from region {\sf I} to region {\sf II}}.
Indeed,
at the interface $x=0$ (for all $y$), the continuity
of the system gives
\begin{eqnarray}
\phi_{{\sf I},+}+r^{+}_{nm}\phi_{{\sf I},-}=t^{+}_{nm}\phi_{{\sf II},+}
\end{eqnarray}
where $r^{+}_{nm}$ and $t^{+}_{nm}$ are reflexion and transmission
coefficients, respectively,  in positive propagation.
From (\ref{7}) and (\ref{8}), we obtain
\begin{eqnarray}
\left(%
\begin{array}{c}
  -si D_{|n|-1}(x_{01}) \\
 D_{|n|}(x_{01}) \\
\end{array}%
\right)+r^{+}_{nm}\left(%
\begin{array}{c}
  -si D_{|n|-1}(-x_{01}) \\
 D_{|n|}(-x_{01}) \\
\end{array}%
\right)=t^{+}_{nm}\left(%
\begin{array}{c}
  -s'i D_{|m|-1}(x_{02}) \\
 D_{|m|}(x_{02}) \\
\end{array}%
\right).
\end{eqnarray}
They can be solved to  get the coefficients in terms of
some constants, which are magnetic field dependent. They are
\begin{eqnarray}
r^{+}_{nm}(x_{01},x_{02})&=&(-1)^{|n|}\frac{sA_{\sf I}-s'B_{\sf I}}
{sA_{\sf I}+s'B_{\sf I}}\\
%\end{eqnarray}
%\begin{eqnarray}
t^{+}_{nm}(x_{01},x_{02})&=& \frac{2sC_{\sf I}}{sA_{\sf I}+s'B_{\sf I}}
\end{eqnarray}
where we have set
\begin{eqnarray}
A_{\sf I} &=& A_{nm}(x_{01},x_{02})=D_{|n|-1}(x_{01})D_{|m|}(x_{02}),\nonumber\\
B_{\sf I} &=& B_{nm}(x_{01},x_{02})=D_{|m|-1}(x_{02})D_{|n|}(x_{01})\nonumber\\
%\end{eqnarray*}
%\begin{eqnarray}
C_{\sf I} &=& C_{n}(x_{01})=D_{|n|-1}(x_{01})D_{|n|}(x_{01})\nonumber.
\end{eqnarray}

On the other hand, considering {propagation  from region {\sf II} to region {\sf I}},
the continuity at point zero reads as
\begin{eqnarray}
\phi_{{\sf II},+}+r^{+}_{mn}\phi_{{\sf II},-}=t^{+}_{mn}\phi_{{\sf I},+}
\end{eqnarray}
which implies
\begin{eqnarray}
\left(%
\begin{array}{c}
  s'i D_{|m|-1}(x_{02}) \\
 D_{|m|}(x_{02}) \\
\end{array}%
\right)+r^{+}_{mn}\left(%
\begin{array}{c}
  s' D_{|m|-1}(-x_{02}) \\
 D_{|m|}(-x_{02}) \\
\end{array}%
\right)=t^{+}_{mn}\left(%
\begin{array}{c}
  s D_{|n|-1}(x_{01}) \\
 D_{|n|}(x_{01}) \\
\end{array}%
\right).
\end{eqnarray}
These lead to the solution
\begin{eqnarray}
r^{+}_{mn}(x_{01},x_{02}) &=& (-1)^{|m|}\frac{s'B_{\sf I}-sA_{\sf I}}{sA_{\sf I}+s'B_{\sf I}+}\\
%\end{eqnarray}
%\begin{eqnarray}
t^{+}_{mn}(x_{01},x_{02}) &=& \frac{2s'F_{\sf I}}{sA_{\sf I}+s'B_{\sf I}}
\end{eqnarray}
where $F_{\sf I}$ is
\begin{eqnarray}
F_{\sf I}=F_{m}(x_{02})=D_{|m|-1}(x_{02})D_{|m|}(x_{02}).
\end{eqnarray}

In similar way, we show that
 the reflexion and transmission coefficients corresponding to
propagation from  {\sf II} to  {\sf III} are given by
\begin{eqnarray}
r^{+}_{mn}(x_{02},x_{03})&=&(-1)^{|m|}\frac{s'B_{\sf II}-sA_{\sf II}}
{sA_{\sf II}+s'B_{\sf II}}\\
t^{+}_{mn}(x_{02},x_{03})&=&\frac{2s'F_{\sf I}}{sA_{\sf II}+s'B_{\sf II}}
\end{eqnarray}
where the constants read as
\begin{equation}
A_{\sf II}=A_{nm}(x_{02},x_{03}),\qquad B_{\sf II}=B_{nm}(x_{02},x_{03}).
\end{equation}
As far as the propagation from  {\sf III} to  {\sf II} is concerned, we find
\begin{eqnarray}
r^{+}_{nm}(x_{02},x_{03}) &=& (-1)^{|m|}\frac{sA_{\sf II}-s'B_{\sf II}}
{sA_{\sf II}+s'B_{\sf II}}\\
%\end{eqnarray}
%\begin{eqnarray}
t^{+}_{nm}(x_{02},x_{03}) &=& \frac{2sC_{\sf II}}{sA_{\sf II}+s'B_{\sf II}}
\end{eqnarray}
where $C_{\sf II}$ is
\begin{eqnarray}
C_{\sf II}=C_{n}(x_{03}).
\end{eqnarray}
This summarizes our analysis for propagation with positive
incidence, which together will be used to discuss different issues
and before doing so, we need to analyze negative incidence.

%%%%%%%%%%%%%%%%%%%%%%%%%%%%%%%%%%%%%%%%%%%%%%%%%%%%%%%
\subsection{Propagation with negative incidence}
%%%%%%%%%%%%%%%%%%%%%%%%%%%%%%%%%%%%%%%%%%%%%%%%%%%%%%%

We determine the reflection and transmission
coefficients for the propagation with negative incidence, which
will be denoted as $r^{-}_{nm/mn}$ and $t^{-}_{nm/mn}$ for the
cases of the propagations from  {\sf I} to {\sf II}, {\sf II} to
{\sf III}  and vice verse. To reply this inquiry, we use the same
analysis as before but one should take into account the negative
sign of variable.

In doing our task, for region {\sf I} we write the corresponding
eigenspinors as
\begin{eqnarray}\label{10}
\phi_{{\sf I},+}=\frac{1}{\sqrt{2}}\left(%
\begin{array}{c}
  -si D_{|n|-1}(-x-x_{01}) \\
 D_{|n|}(-x-x_{01}) \\
\end{array}%
\right)e^{ik_{y}y}, \qquad
\phi_{{\sf I},-}=\frac{1}{\sqrt{2}}\left(%
\begin{array}{c}
  -si D_{|n|-1}(x+x_{01}) \\
 D_{|n|}(x+x_{01}) \\
\end{array}%
\right)e^{ik_{y}y}.
\end{eqnarray}
For region {\sf II}, we have
\begin{eqnarray}\label{11}
\phi_{{\sf II},+}=\frac{1}{\sqrt{2}}\left(%
\begin{array}{c}
  -s'i D_{|m|-1}(-x-x_{02}) \\
       D_{|m|}(-x-x_{02}) \\
\end{array}%
\right)e^{ik_{y}y}, \qquad
\phi_{{\sf II},-}=\frac{1}{\sqrt{2}}\left(%
\begin{array}{c}
  -s'i D_{|m|-1}(x+x_{02}) \\
       D_{|m|}(x+x_{02}) \\
\end{array}%
\right)e^{ik_{y}y}.
\end{eqnarray}
In region {\sf III}, we write
\begin{eqnarray}
\phi_{{\sf III},+}=\frac{1}{\sqrt{2}}\left(%
\begin{array}{c}
  -si D_{|n|-1}(-x-x_{03}) \\
 D_{|n|}(-x-x_{03}) \\
\end{array}%
\right)e^{ik_{y}y}, \qquad
\phi_{{\sf III},-}=\frac{1}{\sqrt{2}}\left(%
\begin{array}{c}
  -si D_{|n|-1}(x+x_{03}) \\
 D_{|n|}(x+x_{03}) \\
\end{array}\
\right)e^{ik_{y}y}.
\end{eqnarray}

Now let us  treat each case to evaluate reflexion and transmission coefficients.
Considering
the propagation: {{\sf I} $\longrightarrow$ {\sf II}} to obtain
\begin{eqnarray}\label{12}
\phi_{{\sf I},+}+r^{-}_{nm}\phi_{{\sf I},-}=t^{-}_{nm}\phi_{{\sf II},+}
\end{eqnarray}
at point zero. Injecting (\ref{10}) and (\ref{11}) into (\ref{12}), one gets
\begin{eqnarray}
r^{-}_{nm}(x_{01},x_{02}) &=& (-1)^{|n|}\frac{sA_{\sf I}-s'B_{\sf I}}{sA_{\sf I}+s'B_{\sf I}}\\
t^{-}_{nm}(x_{01},x_{02}) &=& (-1)^{|n|+|m|}\frac{2sC_{\sf I}}{sA_{\sf I}+s'B_{\sf I}}.
\end{eqnarray}

%%%%%%%%%%%%%%%%%%%%%%%%%%%%%%%%%%%%%%%%%%%%%%%%%%%%%%%%%%%%%%%%
%\subsubsection{{\sf II} $\longrightarrow$ {\sf I}}
%%%%%%%%%%%%%%%%%%%%%%%%%%%%%%%%%%%%%%%%%%%%%%%%%%%%%%%%%%%%%%%%
For {{\sf II} $\longrightarrow$ {\sf I}} a similar relation
to (\ref{12}) reads as
\begin{eqnarray}
\phi_{{\sf II},+}+r^{-}_{mn}\phi_{{\sf II},-}=t^{-}_{mn}\phi_{{\sf I},+}.
\end{eqnarray}
The solutions are given by
\begin{eqnarray}
r^{-}_{mn}(x_{01},x_{02}) &=& (-1)^{|m|}\frac{s'B_{\sf I}-sA_{\sf I}}{sA_{\sf I}+s'B_{\sf I}}\\
t^{-}_{mn}(x_{01},x_{02}) &=& (-1)^{|n|+|m|}\frac{2s'F_{\sf I}}{sA_{\sf I}+s'B_{\sf I}}.
\end{eqnarray}

After applying the same technique as before,
we show that the propagation: {\sf II}  $\longrightarrow$
{\sf III} gives
\begin{eqnarray}
r^{-}_{mn}(x_{02},x_{03}) &=& (-1)^{|m|}\frac{s'B_{\sf II}-sA_{\sf II}}
{sA_{\sf II}+s'B_{\sf II}}\\
t^{-}_{mn}(x_{02},x_{03}) &=& (-1)^{|n|+|m|}\frac{2s'F_{\sf I}}{sA_{\sf II}+s'B_{\sf II}}
\end{eqnarray}
and  {\sf III}  $\longrightarrow$ {\sf II} leads
\begin{eqnarray}
r^{-}_{nm}(x_{02},x_{03}) &=& (-1)^{|n|}\frac{sA_{\sf II}-s'B_{\sf II}}
{sA_{\sf II}+s'B_{\sf II}}\\
t^{-}_{nm}(x_{02},x_{03}) &=& (-1)^{|n|+|m|}\frac{2sC_{\sf II}}{sA_{\sf II}+s'B_{\sf II}}.
\end{eqnarray}

By inspecting the forms of different coefficients obtained so far, one can establish
a symmetry between them. Indeed,
%Note that, we can establish interesting relations between different coefficients
%entering in the game.
We show the relation
\begin{equation}
%r_{ij}^{\pm}(x_{01},x_{02}) r_{ij}^{\mp}(x_{01},x_{02}) , \qquad
\frac{t_{ij}^{\pm}(x_{01},x_{02})}{t_{ij}^{\mp}(x_{01},x_{02})}
=-\frac{r_{ij}^{\pm}(x_{01},x_{02})}{r_{ji}^{\pm}(x_{01},x_{02})}=(-1)^{|n|+|m|}
\end{equation}
where the pair of index is chosen  such as $(i \neq j)$ $\in \{n,m\}$. Note that,
the same relations are also valid for the couple
$(x_{02},x_{03})$.

%%%%%%%%%%%%%%%%%%%%%%%%%%%%%%%%%%%%%%%%%%%%%%%%%%%%%%%%%%%%%%%%%
\subsection{ Reflexion and transmission amplitudes}
%%%%%%%%%%%%%%%%%%%%%%%%%%%%%%%%%%%%%%%%%%%%%%%%%%%%%%%%%%%%%%%%%%

Now let us collect the products of our results by checking their
importance. In fact, we discuss the reflexion and transmission amplitudes
between regions to emphasis the influence of each parameter
on them.
%(\textcolor[rgb]{0.00,0.00,1.00}{et aussi  discuter la
%transmission et la r\'{e}flexion entres les r\'{e}gions et  voire
%qu'il est l'impacte de chaque param\`{e}tres sur la transmission
%et la r\'{e}flexion}).
% defining different coefficients
%in terms of what we obtained so far.
%For this, we consider the first
For propagation between {\sf I} and {\sf II}, we define the
reflexion and transmission amplitudes as
\begin{equation}
\rho(x_{01},x_{02})=r_{ij}^{\pm}(x_{01},x_{02})r_{ij}^{\mp}(x_{01},x_{02}),
\qquad
\tau(x_{01},x_{02})=t_{ij}^{\pm}(x_{01},x_{02})t_{ji}^{\pm}(x_{01},x_{02}).
\end{equation}
After replacing, we end up with
\begin{eqnarray}
\rho(x_{01},x_{02}) &=&\frac{\left[sA_{\sf I}-s'B_{\sf
I}\right]^{2}}
{\left[sA_{\sf I}+s'B_{\sf I}\right]^2}\label{133}\\
\tau(x_{01},x_{02}) &=&\frac{4ss'C_{\sf I}F_{\sf I}}
{\left[sA_{\sf I}+s'B_{\sf I}\right]^{2}}=\frac{4ss'A_{\sf
I}B_{\sf I}} {\left[sA_{\sf I}+s'B_{\sf I}\right]^2}\label{134}
\end{eqnarray}
where the relation $C_{\sf I}F_{\sf I}=A_{\sf I}B_{\sf I}$ is satisfied.
A straightforward calculation shows that the probability sums to unity,
namely
\begin{equation}\label{13}
\rho(x_{01},x_{02})+\tau(x_{01},x_{02})=1.
\end{equation}
%\textcolor[rgb]{0.00,0.00,1.00}{Lesse nous maintenant discuter la
%transmission et la r\'{e}flexion entres les r\'{e}gions et aussi
%voire qu'il est l'impacte de chaque param\`{e}tres sur la
%transmission et la r\'{e}flexion.}
To stress how the amplitudes behave in terms of different parameters,
we give Figure 5. It is clear that, %the reflexion and transmission
$\rho(x_{01},x_{02})$ and $\tau(x_{01},x_{02})$
change with respect to variation the magnetic field $B'$ for
different values of $B,m,d$ and $k_y$.

%%%%%%%%%%%%%%%%%%%%%%%%%%%%%%%%%%%%%%%%%%%%%%%%%%%%%%%%%%%%%%%%%%%%%%
\begin{center}
  \includegraphics[width=6.7in]{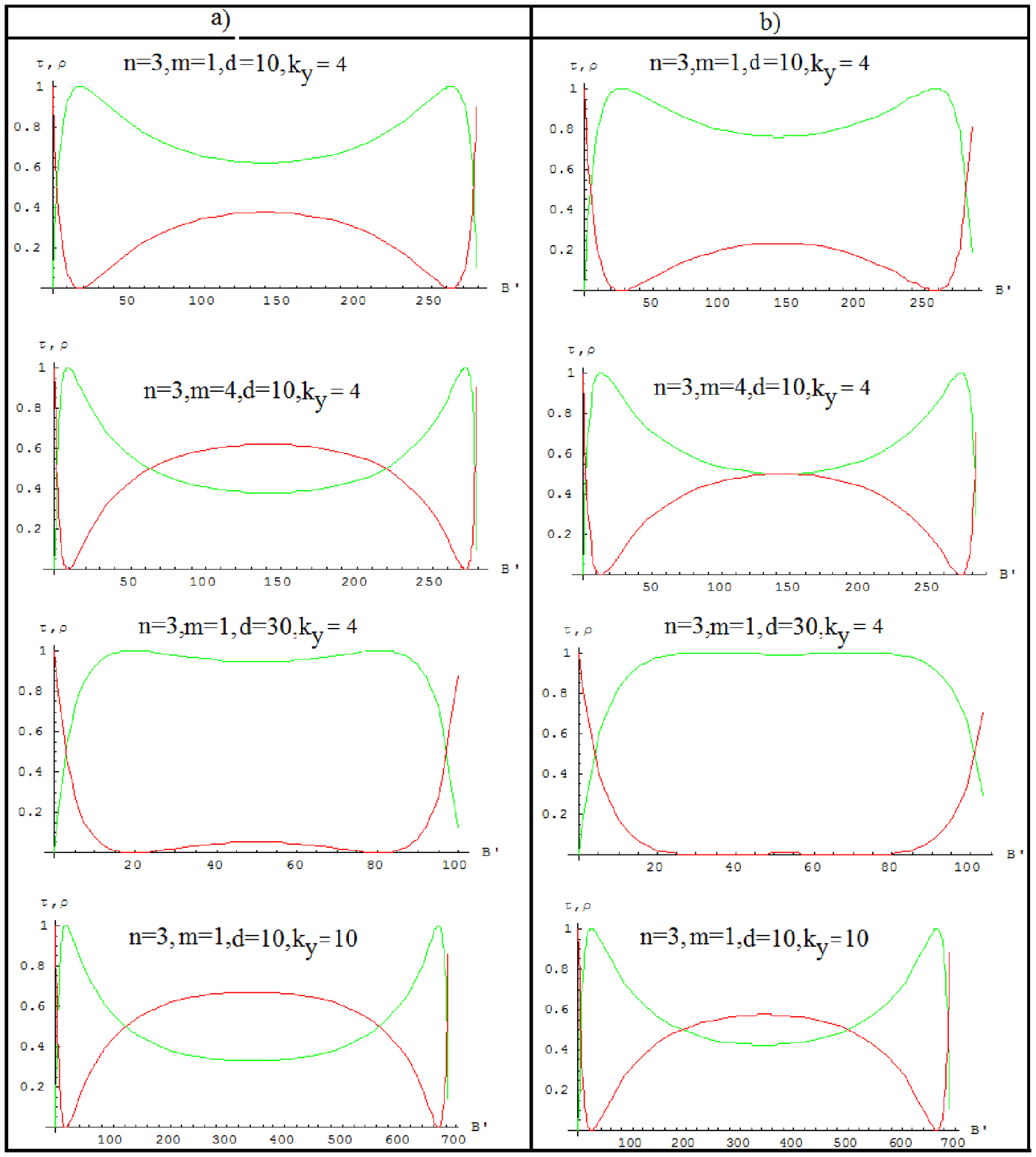}\\
\end{center}
{\sf Figure 5:} {{ Reflexion $\rho(x_{01},x_{02})$ (red line) and
transmission $\tau(x_{01},x_{02})$ (green line) coefficients
between region {\sf I} and {\sf II} for a magnetic barrier of
width $2d$ at various magnetic field $B'$ for two cases:  a)
$B=10$,  b) $B=15$ and for different values of $(m, d, k_{y})$}}.\\

%%%%%%%%%%%%%%%%%%%%%%%%%%%%%%%%%%%%%%%%%%%%%%%%%%%%%%%%%%%%%%%%%%%%

\noindent It worthwhile to invert the situation by varying  $B$ for
two values of $B'$ a given configuration of the parameters $(m, d, k_{y})$.
This summarized as follows

%%%%%%%%%%%%%%%%%%%%%%%%%%%%%%%%%%%%%%%%%%%%%%%%%%%%%%%%%%%%%%%%%%%%%%
\begin{center}
  \includegraphics[width=6.7in]{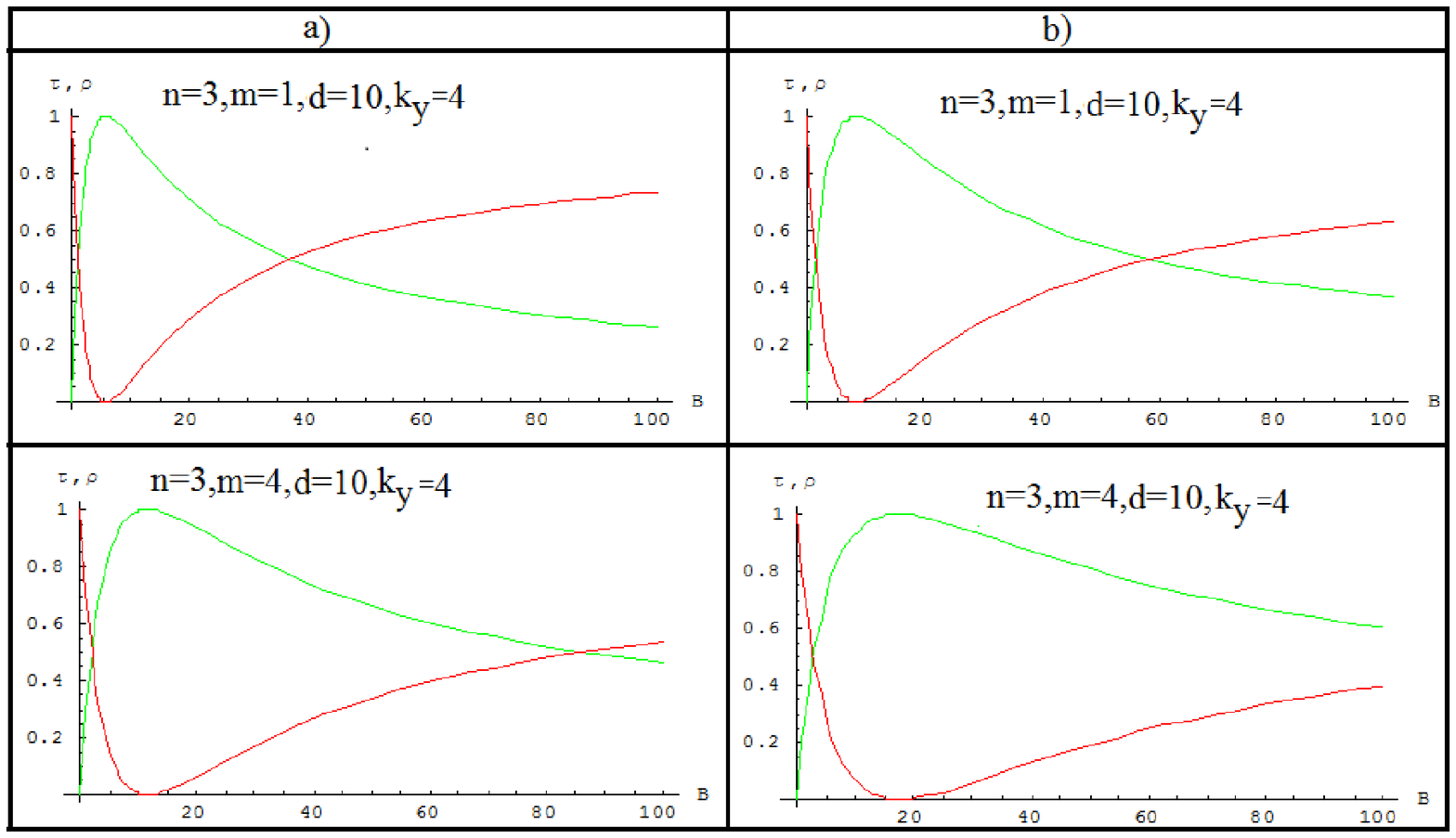}\\
\end{center}
{\sf Figure 6:} {{ Reflexion $\rho(x_{01},x_{02})$ (red line) and
transmission $\tau(x_{01},x_{02})$ (green line) coefficients
between region {\sf I} and {\sf II} for a magnetic barrier of
width $2d$ at various magnetic field $B$ for two cases:  a)
$B'=10$, b) $B'=15$ and for different values of $(m, d, k_{y})$}}.\\
%%%%%%%%%%%%%%%%%%%%%%%%%%%%%%%%%%%%%%%%%%%%%%%%%%%%%%%%%%%%%%%%%%%%

As far as the propagation
%Now let us inspect the second case, which is
between {\sf II} and {\sf III} is concerned, we use the same
definition as above to write the amplitudes
\begin{equation}
\rho(x_{02},x_{03})=r_{ij}^{\pm}(x_{02},x_{03})r_{ij}^{\mp}(x_{02},x_{03}),
\qquad
\tau(x_{02},x_{03})=t_{ij}^{\pm}(x_{02},x_{03})t_{ji}^{\pm}(x_{02},x_{03})
\end{equation}
which give
\begin{eqnarray}
\rho(x_{02},x_{03}) &=&\frac{\left[sA_{\sf II}-s'B_{\sf II}\right]^{2}}
{\left[sA_{\sf II}+s'B_{\sf II}\right]^2}\\
\tau(x_{02},x_{03}) &=&\frac{4ss'C_{\sf II}F_{\sf I}}
{\left[sA_{\sf II}+s'B_{\sf II}\right]^{2}}=\frac{4ss'A_{\sf II}B_{\sf II}}
{\left[sA_{\sf II}+s'B_{\sf II}\right]^2}
\end{eqnarray}
where  $C_{\sf II}F_{\sf I}=A_{\sf II}B_{\sf II}$. Using these
to verify
\begin{equation}\label{14}
\rho(x_{02},x_{03})+\tau(x_{02},x_{03})=1.
\end{equation}
As before we illustrate this result by making different plots, which are
given by Figure 7

%%%%%%%%%%%%%%%%%%%%%%%%%%%%%%%%%%%%%%%%%%%%%%%%%%%%%%%%%%%%%%%%%%%%%%
\begin{center}
  \includegraphics[width=6.7in]{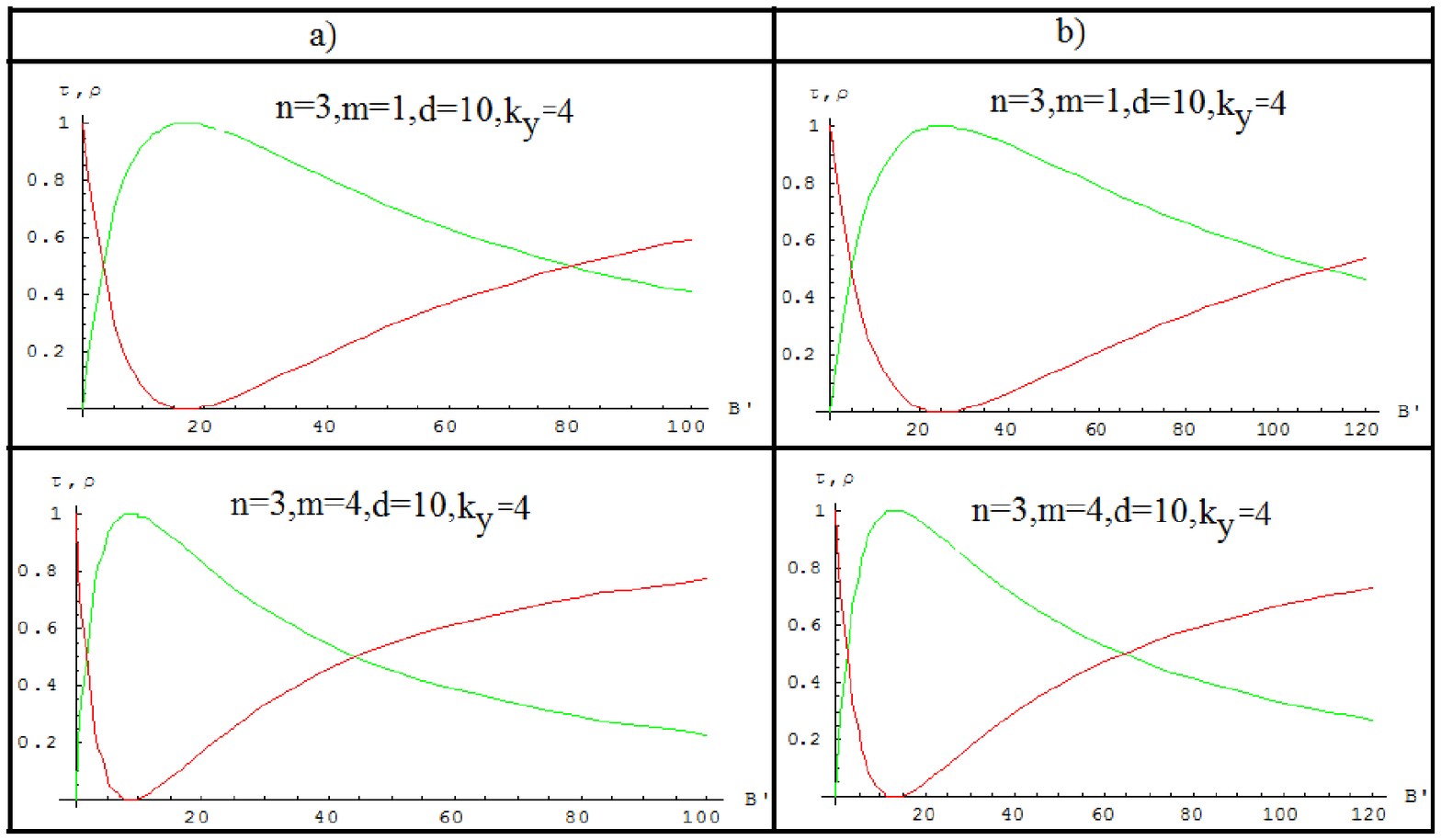}\\
\end{center}
{\sf Figure 7:} {{ Reflexion $\rho(x_{02},x_{03})$ (red line) and
transmission $\tau(x_{02},x_{03})$ (green line) coefficients
between region {\sf II} and {\sf III} for a magnetic barrier of
width $2d$ at various magnetic field $B'$ for two cases:  a)
$B=10$, b) $B=15$ and for different values of $(m, d, k_{y})$}}.

%\end{center}
%%%%%%%%%%%%%%%%%%%%%%%%%%%%%%%%%%%%%%%%%%%%%%%%%%%%%%%%%%%%%%%%%%%%
%%%%%%%%%%%%%%%%%%%%%%%%%%%%%%%%%%%%%%%%%%%%%%%%%%%%%%%%%%%%%%%%%%%%%%
\begin{center}
  \includegraphics[width=6.7in]{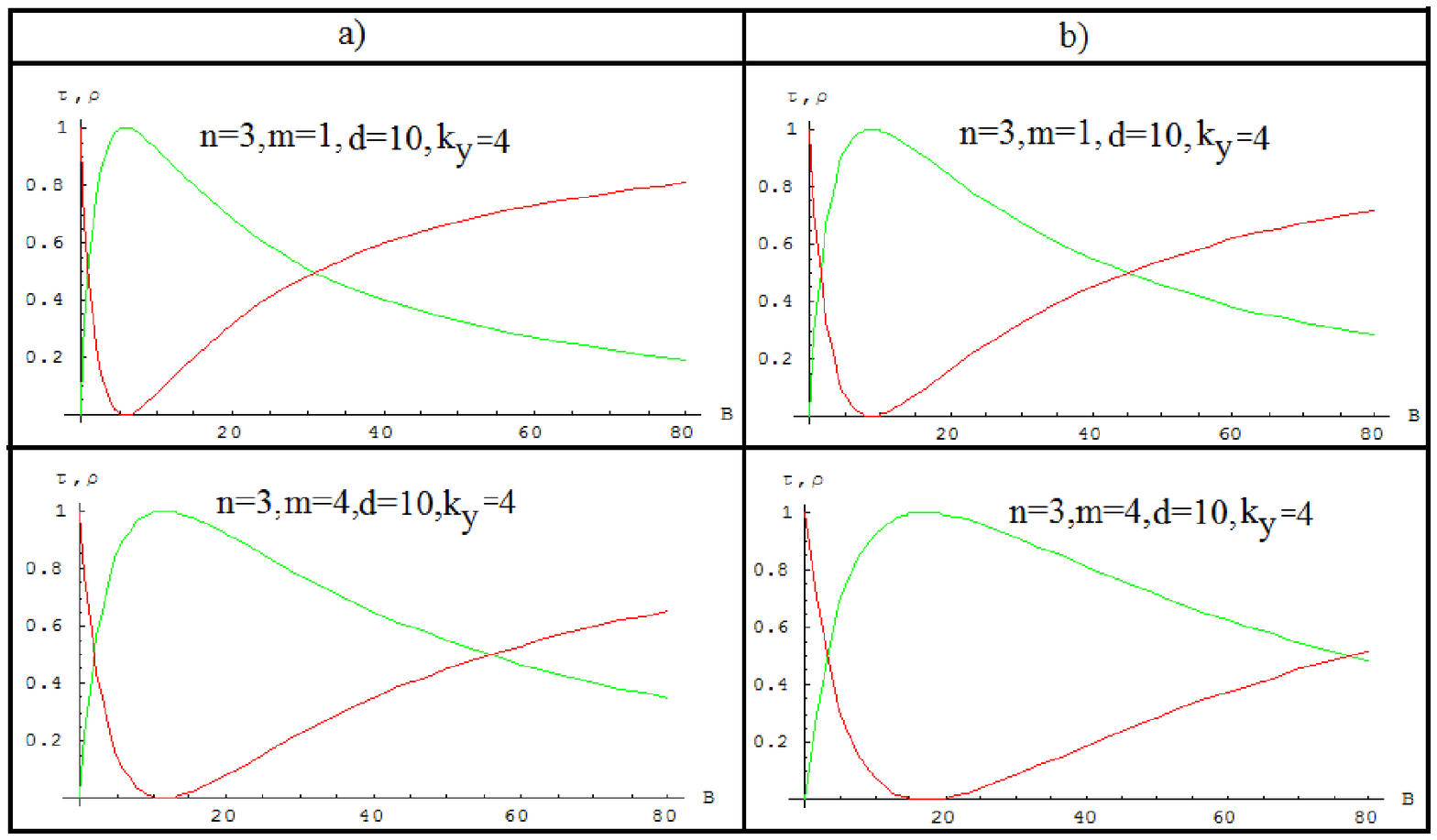}\\
  \end{center}
{\sf Figure 8:} {{ Reflexion $\rho(x_{02},x_{03})$ (red line) and
transmission $\tau(x_{02},x_{03})$ (green line) coefficients
between region {\sf II} and {\sf III} for a magnetic barrier of
width $2d$ at various magnetic field $B$ for two cases:  a)
$B'=10$, b) $B'=15$ and for different values of $(m, d, k_{y})$}}.\\

\noindent
%%%%%%%%%%%%%%%%%%%%%%%%%%%%%%%%%%%%%%%%%%%%%%%%%%%%%%%%%%%%%%%%%%%%
These are among the interesting results derived so far. In fact, it
tell us the transmission of barrier in inhomogeneous magnetic fields can not be
greater than one, which is analogue to what obtained in one field
case and without confinement~\cite{cho}.

%%%%%%%%%%%%%%%%%%%%%%%%%%%%%%%%%%%%%%%%%%%%%%%%%%%%%%%
%\subsection{Discussions}
%%%%%%%%%%%%%%%%%%%%%%%%%%%%%%%%%%%%%%%%%%%%%%%%%%%%%%%

Let us present  some discussions and derive  interesting results those
have applications in physics areas. %give some interesting discussions
%those have  known results.
Indeed,
%It will be very useful in the following, for expression
%simplification purposes, the derivation of simple relations
%between the reflection and transmission amplitude coefficients.
similar relations to what obtained above exist also for the
photon optics cases. For instance, one can write  (\ref{13})
and  (\ref{14}) when a light beam is reflected and refracted in
a diopter between regions {\sf I}-{\sf II} and {\sf II}-{\sf III}.
The first case is characterized by the configuration
\begin{itemize}
    \item For $n=0$ \qquad $\Longrightarrow$ \qquad $\left\{%
\begin{array}{ll}
    \rho(x_{01},x_{02})=1 , \qquad \tau(x_{01},x_{02})=0  \\
    \rho(x_{02},x_{03})=1 , \qquad \tau(x_{02},x_{03})=0.  \\
\end{array}%
\right.$\\
\end{itemize}
This is an expected result since the transmission  $ \tau= 1 -
\rho$ must be zero in the case where the wave in the $n$-region
enters in the $m$-region and vice verse. In fact, the
interface between tree regions behaves like a mirror where the
reflexion is total. The second case is described by

\begin{itemize}
    \item For $s=s'$ and $n=\pm m$ \qquad $\Longrightarrow$ \qquad $\left\{%
\begin{array}{ll}
    \rho(x_{01},x_{02})=0 , \qquad \tau(x_{01},x_{02})=1  \\
    \rho(x_{02},x_{03})=0 , \qquad \tau(x_{02},x_{03})=1  \\
\end{array}%
\right.$\\
\end{itemize}
which means that the interface between region {\sf I}-{\sf II} and {\sf II}-{\sf III} behaves
like a non-reflective diopter, namely there is a total transmission. Note that, these two cases have
interesting interpretation in optics physics.

%%%%%%%%%%%%%%%%%%%%%%%%%%%%%%%%%%%%%%%%%%%%%%%%%%%%%%%%%%%%%%%%%%%%%%
\section{Tree regions in inhomogeneous magnetic fields}
%%%%%%%%%%%%%%%%%%%%%%%%%%%%%%%%%%%%%%%%%%%%%%%%%%%%%%%%%%%%%%%%%%%%%%

We study another case of
%Let us consider
a physical system composed of a region indexed by the
quantum number $m$ of length $2w$ separating two others indexed by the
same quantum number $n$. This will allow us
to see how the above results will be changed
to the present case and underline what make difference with respect to the
former analysis.
% will apply the machinery as before to
%analyze the system behavior in such configuration.

%%%%%%%%%%%%%%%%%%%%%%%%%%%%%%%%%%%%%%%%%%%%%%%%%%%%%%%%%%%%%%%%%%%%%%
\subsection{Reflexion and transmission coefficients}
%%%%%%%%%%%%%%%%%%%%%%%%%%%%%%%%%%%%%%%%%%%%%%%%%%%%%%%%%%%%%%%%%%%%%%

The present situation is quiet different from the former one.
We use the above tool to write the continuity equation at
the points $x=-d$ and $x=d$. Then, we derive the
quantities needed to discuss the reflexion and transmission
coefficients as well as the corresponding probabilities. This will be done
by %As before
%we separately
treating
propagation with positive and negative incidences.

We study the positive incidence by
considering the geometry that corresponds to
the first interface, which is

%%%%%%%%%%%%%%%%%%%%%%%%%%%%%%%%%%%%%%%%%%%%%%%%%%%%%%%%%%%%%%%%%%%%%%
\begin{center}
  \includegraphics[width=2.5in]{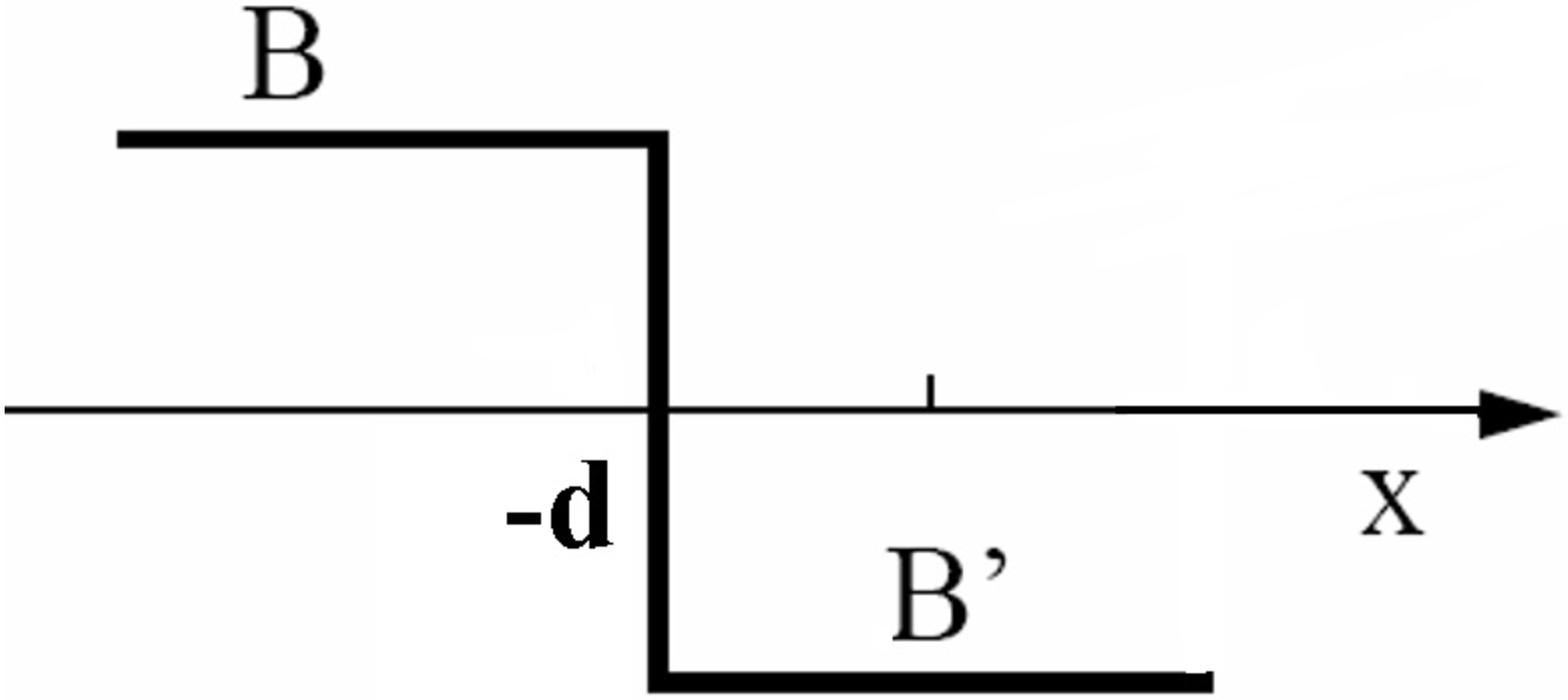}\\
{\sf Figure 9:} {{Magnetic field profile between regions {\sf I}
and {\sf II}}}.
\end{center}
%%%%%%%%%%%%%%%%%%%%%%%%%%%%%%%%%%%%%%%%%%%%%%%%%%%%%%%%%%%%%%%%%%%%
Using the above spinors to setup the continuity equation at
the point $x=-d$. Doing this process to get the relation
\begin{equation}\label{poinc}
\phi_{{\sf I},+}(-d)+r^{+}\phi_{{\sf I},-}(-d)=\alpha\phi_{{\sf II},+}(-d)+\beta\phi_{{\sf II},-}(-d)
\end{equation}
where $r^{+}$ is the reflection coefficient for positive incidence,
which will be determined together with
 the parameters $\alpha$ and $\beta$.  (\ref{poinc}) gives
\begin{eqnarray}
 && sD_{|n|-1}(d_{1})+r^{+}sD_{|n|-1}(-d_{1}) = \alpha
s'D_{|m|-1}(d_{2}) +\beta s'D_{|m|-1}(-d_{2})\\
 && D_{|n|}(d_{1})+r^{+}D_{|n|}(-d_{1}) = \alpha D_{|m|}(d_{2}) +\beta
D_{|m|}(-d_{2})
\end{eqnarray}
with the constants $d_{1}=x_{01}-d$ and $d_{2}=x_{02}-d$. These can be solved
for $\al$ and $\be$ to obtain
\begin{eqnarray}
\alpha &=& \frac{sA_{nm}(d_{1},d_{2})+s'B_{nm}(d_{1},d_{2})
+r^{+}(-1)^{|n|}\left[s'B_{nm}(d_{1},d_{2})
-sA_{nm}(d_{1},d_{2})\right]}{2s'F_{m}(d_{2})}\\
\beta &=& \frac{s'B_{nm}(d_{1},d_{2})-sA_{nm}(d_{1},d_{2})
+r^{+}(-1)^{|n|}\left[sA_{nm}(d_{1},d_{2})+s'B_{nm}(d_{1},d_{2})
\right]}{2(-1)^{|m|}s'F_{m}(d_{2})}.
\end{eqnarray}
In terms of the reflexion and transmission coefficients,
we have
\begin{eqnarray}
\alpha &=& \frac{1}{t^{+}_{mn}(d_{1},d_{2})}
-r^{+}\frac{r^{-}_{nm}(d_{1},d_{2})}{t^{+}_{mn}(d_{1},d_{2})}\label{d21}\\
\beta &=&  (-1)^{|n|+|m|}\left[\frac{r^{+}}{t^{+}_{mn}(d_{1},d_{2})}
-\frac{r^{-}_{nm}(d_{1},d_{2})}{t^{+}_{mn}(d_{1},d_{2})}\right]\label{d22}.
\end{eqnarray}

To accomplish such analysis we consider the second interface as shown below
%%%%%%%%%%%%%%%%%%%%%%%%%%%%%%%%%%%%%%%%%%%%%%%%%%%%%%%%%%%%%%%%%%%%%%%%%%%%%%%%%
\begin{center}
  \includegraphics[width=2in]{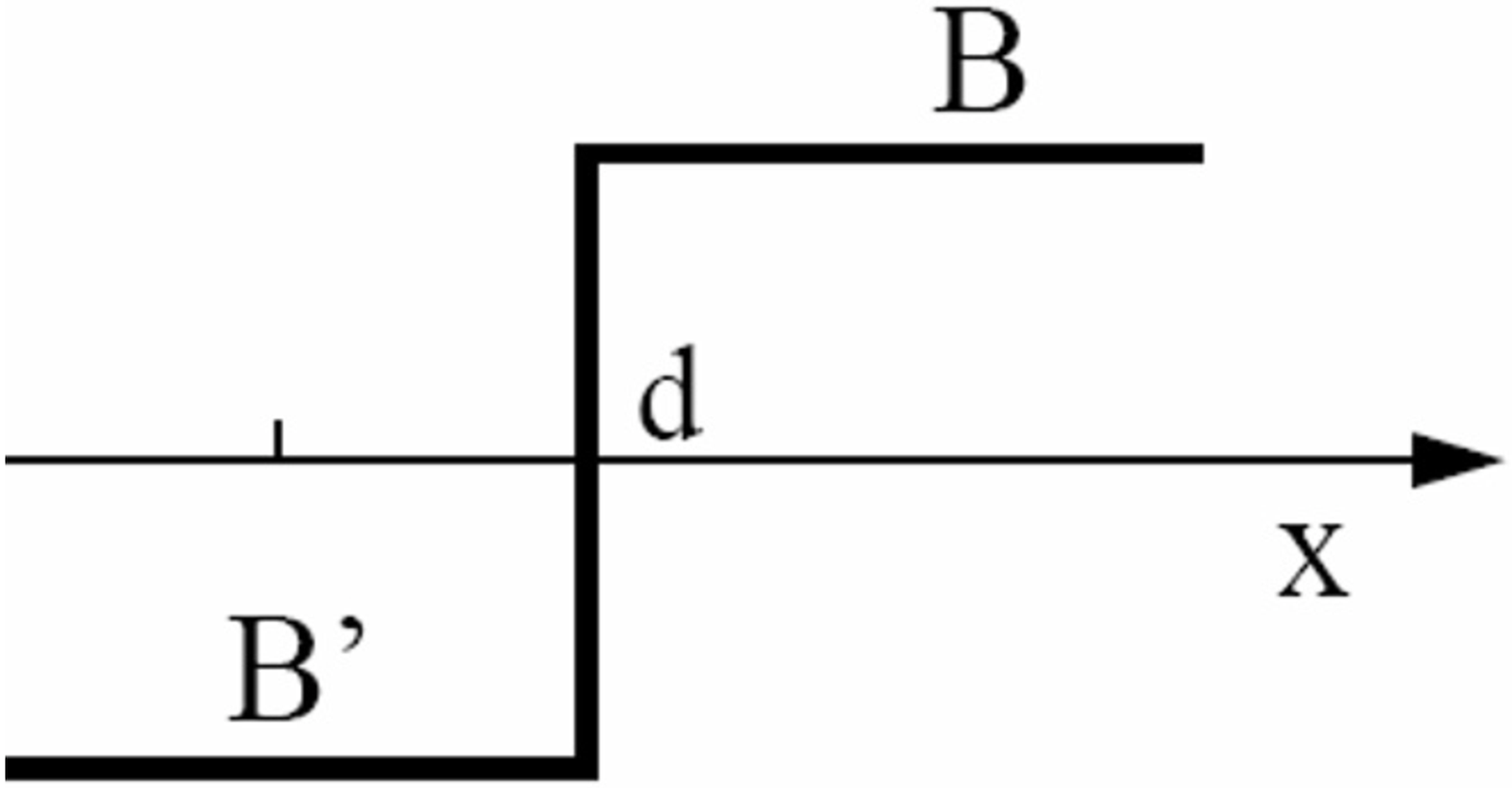}\\
{\sf Figure 10:}  {{Magnetic field profile between region {\sf II}
and {\sf III$\equiv$I}}}.
\end{center}
%%%%%%%%%%%%%%%%%%%%%%%%%%%%%%%%%%%%%%%%%%%%%%%%%%%%%%%%%%%%%%%%%%%%%%%%%%%%%%%%
At the point $x=d$, we have
\begin{eqnarray}
\alpha\phi_{{\sf II},+}(d)+\beta\phi_{{\sf II},-}(d)=t^{+}\phi_{{\sf III},+}(d)
\end{eqnarray}
where $t^{+}$ is transmission coefficient for positive incidence. After
replacing, we end up with a system of equations, such as
\begin{eqnarray}
&& \alpha s'D_{|m|-1}(d_{2}') +\beta s'D_{|m|-1}(-d_{2}') = t^{+}sD_{|n|-1}(d_{3})\\
&& \alpha D_{|m|}(d_{2}') +\beta D_{|m|}(-d_{2}') = t^{+}D_{|n|}(d_{3})
\end{eqnarray}
with $d_{2}'=d+x_{02}$ and $d_{3}=d+x_{03}$. The solution reads as
\begin{eqnarray}
\alpha &=& t^{+}\frac{sA_{nm}(d_{3},d_{2}')+s'B_{nm}(d_{3},d_{2}')}{2s'F_{m}(d_{2}')}\nonumber\\
&=&\frac{t^{+}}{t^{+}_{mn}(d_{2}',d_{3})}\label{d2pr1}\\
\beta &=& t^{+}(-1)^{|m|}\frac{s'B_{nm}(d_{3},d_{2}')-sA_{nm}(d_{3},d_{2}')}
{2s'F_{m}(d_{2}')}\nonumber\\
&=& -t^{+}(-1)^{|n|+|m|}\frac{r^{-}_{n,m}(d_{2}',d_{3})}{t^{+}_{mn}(d_{2}',d_{3})}\label{d2pr2}
\end{eqnarray}
To determine the coefficients for positive incidence
we simply use  (\ref{d21}-\ref{d22}) and (\ref{d2pr1}-\ref{d2pr2}). Combing all
to obtain
\begin{eqnarray}
r^{+} &=&  \frac{r^{-}_{nm}(d_{1},d_{2})-r^{-}_{nm}(d_{2}',d_{3})}
{1-r^{-}_{nm}(d_{1},d_{2})r^{-}_{nm}(d_{2}',d_{3})}\\
t^{+} &=& \frac{t^{+}_{mn}(d_{2}',d_{3})}{t^{+}_{mn}(d_{1},d_{2})}\left[
\frac{1-r^{-}_{nm}(d_{1},d_{2})r^{-}_{nm}(d_{1},d_{2})}{1-r^{-}_{nm}
(d_{1},d_{2})r^{-}_{nm}(d_{2}',d_{3})}\right]\nonumber\\
&=&\frac{t^{+}_{mn}(d_{2}',d_{3})}{t^{+}_{mn}(d_{1},d_{2})}\left[
\frac{\tau(d_{1},d_{2})}{1-r^{-}_{nm}(d_{1},d_{2})r^{-}_{nm}(d_{2}',d_{3})}\right].
\end{eqnarray}

Now let see how the above results will be written by considering the negative
incidence case. Indeed,  applying the same machinery as before to get
\begin{eqnarray}
r^{-} &=& \frac{r^{+}_{n,m}(d_{1},d_{2})-r^{+}_{n,m}(d_{2}',d_{3})}
{1-r^{+}_{n,m}(d_{1},d_{2})r^{+}_{n,m}(d_{2}',d_{3})} \\
&=&
\frac{r^{-}_{n,m}(d_{1},d_{2})-r^{-}_{n,m}(d_{2}',d_{3})}
{1-r^{-}_{n,m}(d_{1},d_{2})r^{-}_{n,m}(d_{2}',d_{3})}\nonumber\\
t^{-} &=& \frac{t^{-}_{n,m}(d_{2}',d_{3})}{t^{-}_{n,m}(d_{1},d_{2})}\left[
\frac{\tau(d_{1},d_{2})}{1-r^{+}_{n,m}(d_{1},d_{2})r^{+}_{n,m}(d_{2}',d_{3})}\right]\nonumber\\
&=&\frac{t^{-}_{n,m}(d_{2}',d_{3})}{t^{-}_{n,m}(d_{1},d_{2})}\left[
\frac{\tau(d_{1},d_{2})}{1-r^{-}_{n,m}(d_{1},d_{2})r^{-}_{n,m}(d_{2}',d_{3})}\right].
\end{eqnarray}
Having obtained the above results, we analyze the corresponding probability
and give comments. %one may ask about their relevance. To reply this inquiry,
%we turn back and use some analysis in order to derive different conclusions.
This issue
and related matter will be considered in the forthcoming subsection.

%of the coming subsection.

%%%%%%%%%%%%%%%%%%%%%%%%%%%%%%%%%%%%%%%%%%%%%%%%%%%%
\subsection{Probability}
%%%%%%%%%%%%%%%%%%%%%%%%%%%%%%%%%%%%%%%%%%%%%%%%%%

To characterize the behavior of the present system, we study the incident beam. This can
be achieved by calculating the probability of reflexing and transmitting
beam. Indeed,
%Roughly speaking, one can use the above results to end up with
%interesting results and different conclusions.
%
%To be much more  precise,
%Now let us collect the product of our findings. In doing so,
let us
adopt the definition %for the
%The general definitions for the intensity
%reflection and transmission amplitudes. This allows us to write
\begin{eqnarray}
R=r^{+}r^{-}, \qquad T=t^{+}t^{-}.
\end{eqnarray}
%where we keep the same notation as before.
After calculation, we find
\begin{eqnarray}\label{82}
R &=& \frac{\rho(d_{1},d_{2})+\rho(d_{2}',d_{3})
-2r^{-}_{n,m}(d_{1},d_{2})r^{-}_{n,m}(d_{2}',d_{3})}
{1+\rho(d_{1},d_{2})\rho(d_{2}',d_{3})
-2r^{-}_{n,m}(d_{1},d_{2})r^{-}_{n,m}(d_{2}',d_{3})} \label{82}\\
T &=& \frac{1+\rho(d_{1},d_{2})\rho(d_{2}',d_{3})
-\rho(d_{1},d_{2})-\rho(d_{2}',d_{3})}
{1+\rho(d_{1},d_{2})\rho(d_{2}',d_{3})
-2r^{-}_{n,m}(d_{1},d_{2})r^{-}_{n,m}(d_{2}',d_{3})}\label{83}.
\end{eqnarray}
Combining all to end up with probability
\begin{equation}
R +T =1.
\end{equation}
From this, we summarize the following conclusions:
\begin{itemize}
\item %In other words,
The probabilities of reflection and transmission
%In other words, the probabilities of reflection and transmission
sum to unity, as must be the case, since they are the only possible outcomes for a fermion
incident on the barrier.
%\item This is among the interesting conclusion
%reached in this section.
%Furthermore,
\item(\ref{82}) and (\ref{83}) yield that under resonance
conditions:
 $$|n|=|m|, \qquad s=s'$$
 the barrier becomes transparent, i.e. $T=1$.
\item More significantly, however, the barrier remains always
perfectly transparent for $|n|=|m|$.
\item $T=1$ is the feature
unique to massless Dirac fermions.
\item $T=1$ is directly related to the
Klein paradox in quantum electrodynamics.

\end{itemize}

%%%%%%%%%%%%%%%%%%%%%%%%%%%%%%%%%%%%%%%%%%%%%%%%%%%%%%%
\section{Introducing gap}
%%%%%%%%%%%%%%%%%%%%%%%%%%%%%%%%%%%%%%%%%%%%%%%%%%%%%%%
In the present study, we consider the confined system in inhomogeneous
magnetic field given by the configuration  (1) but in the presence of an energy gap $t'$ in the
region {\sf II}. We will show how the above results will be generalized
to the gap case.
% and we calculate the transmission probabilities of
%massless Dirac particles.

%%%%%%%%%%%%%%%%%%%%%%%%%%%%%%%%%%%%%%%%%%%%%%%%%
\subsection{Hamiltonian formalism}
%%%%%%%%%%%%%%%%%%%%%%%%%%%%%%%%%%%%%%%%%%%

As far as  regions {\sf I} and {\sf III}
are concerned,
the eigenvalues and the
eigenfunctions are those given before for case of without $t'$. However,
in region {\sf II} the Dirac Hamiltonian can be written as
\begin{eqnarray}
H_{\sf II}=H_{q_{y}}^{B'}=\upsilon_{F}\vec{\sigma} \vec{\pi}+V_{0}
+t'\sigma_{z}.
\end{eqnarray}
Clearly, the mass term $t'\sigma_{z}$ makes difference with respect to the former analysis.
In fact, it will play a crucial role and lead to discover interesting results.
%between both the above Hamiltonian's.
In terms of matrix, $H_{\sf II}$ takes the form
%From the nature of the present system, we write the
%Hamiltonian corresponding to region {\sf II} in matrix form as
\begin{eqnarray}\label{hamgap}
H_{\sf II}= \upsilon_{F}\left(%
\begin{array}{cc}
 0 & p_{x}-ip_{y}-i\frac{e}{c}A_{2}(x) \\
  p_{x}+ip_{y}+i\frac{e}{c}A_{2}(x) & 0 \\
\end{array}%
\right)+\left(%
\begin{array}{cc}
  V_{0}+ t' & 0 \\
  0 & V_{0}-t' \\
\end{array}%
\right).
\end{eqnarray}

For next purpose, we determine the energy spectrum solutions of (\ref{hamgap}). In
doing so, let us fix $\phi_{\sf II}=\left(\begin{array}{c}
  \varphi_{1}' \\
  \varphi_{2}' \\
\end{array}\right)$ as a spinor of $H_{\sf II}$ in presence of an energy gap $t'$,
such as
\begin{eqnarray}
H_{\sf II}\left(\begin{array}{c}
  \varphi_{1}' \\
  \varphi_{2}' \\
\end{array}\right)=E_{\sf II}\left(\begin{array}{c}
  \varphi_{1}' \\
  \varphi_{2}' \\
\end{array}\right)
\end{eqnarray}
which implies two relations
\begin{eqnarray}\label{16}
&& -i\hbar\omega_{c}'D_{2}\varphi_{2}'=(E_{\sf II}-V_{0}-t')\varphi_{1}'
\\ \label{17}
&& i\hbar\omega_{c}'D_{2}^{+}\varphi_{1}'=(E_{\sf II}-V_{0}+t')\varphi_{2}'.
\end{eqnarray}
%(\ref{16}) into (\ref{17}), we obtain
They are showing
\begin{eqnarray}\label{18}
\hbar^{2}\omega_{c}'^{2}D_{2}^{+}D_{2}\varphi_{2}'=\left[(E_{II}+V_0)^{2}-t'^{2}\right]\varphi_{2}'.
\end{eqnarray}
It solution gives the second spinor component as
\begin{eqnarray}
\varphi_{2}'(x,y)=D_{|m|}\left(x+x_{02}\right)e^{iq_{y}y} , \qquad
m\in \mathbb{Z}.
\end{eqnarray}
From (\ref{18}), it is easy to obtain the eigenvalues
\begin{eqnarray}\label{ah}
E_{{\sf II},m}=s'\sqrt{\hbar^{2}\omega_{c}'^{2}|m|+t'^{2}}+V_{0}.
\end{eqnarray}
%It is amusing to notice that the spectrum (\ref{ah}) has the same
%form as for diode in zero field~\cite{ppp}.
%%%%%%%%%%%%%%%%%%%%%figure%%%%%%%%%%%%%%%%%%%%%%%%%%%%%%
\begin{center}
% \begin{figure}
 \includegraphics[width=2.2in]{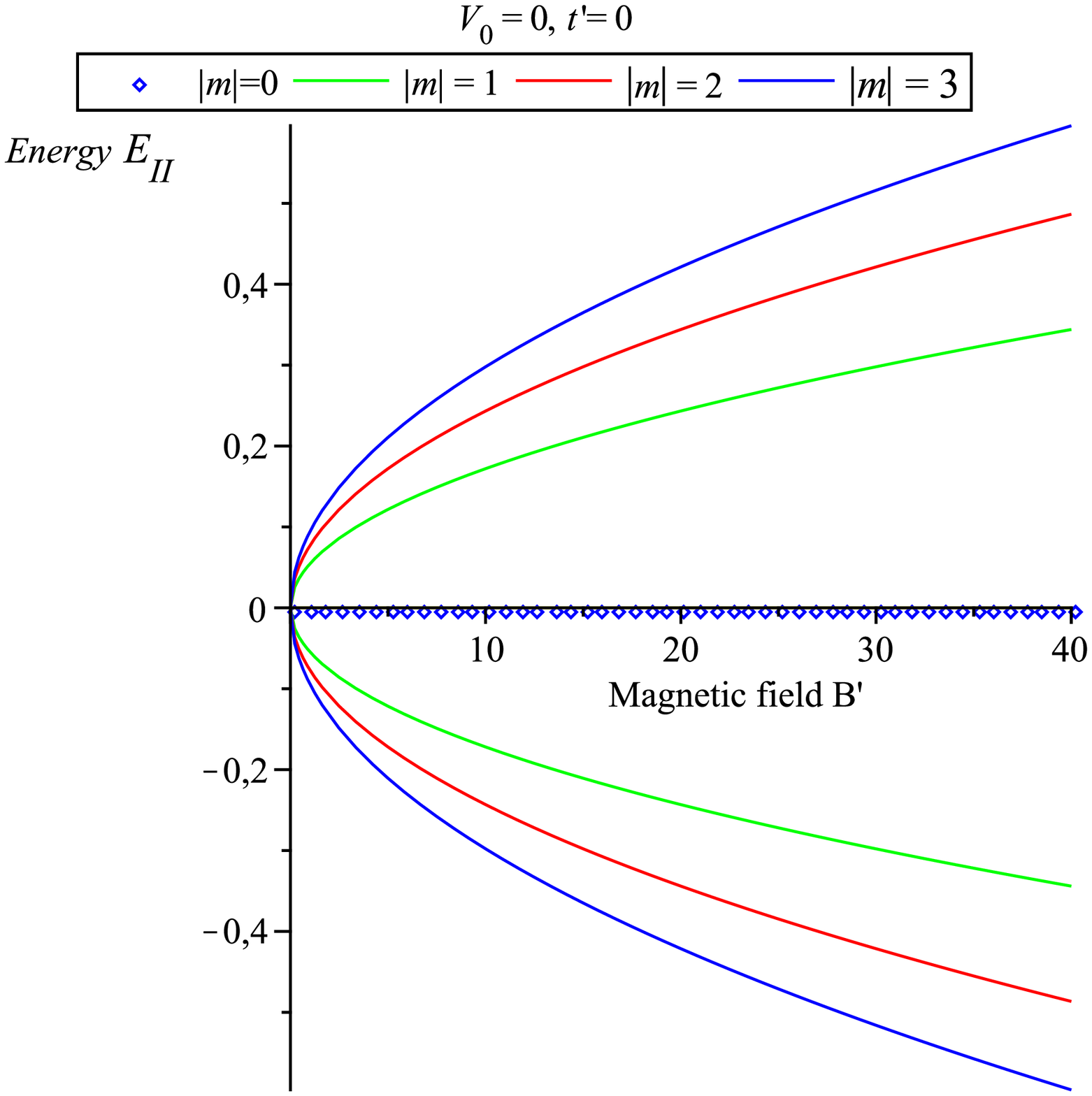}\\
 \end{center}
 {\sf Figure 11}: Landau levels $E_{m}$ in terms of magnetic field  $B'$ for different values of $m$.
%\end{figure}
%\end{center}
%%%%%%%%%%%%%%%%%%%%%%%%%%%%%%%%%%%%%%%%%%%%%%%%%%%%%%%%%
%%%%%%%%%%%%%%%%%%%%%figure%%%%%%%%%%%%%%%%%%%%%%%%%%%%%%
\begin{center}
% \begin{figure}
 \includegraphics[width=2.5in]{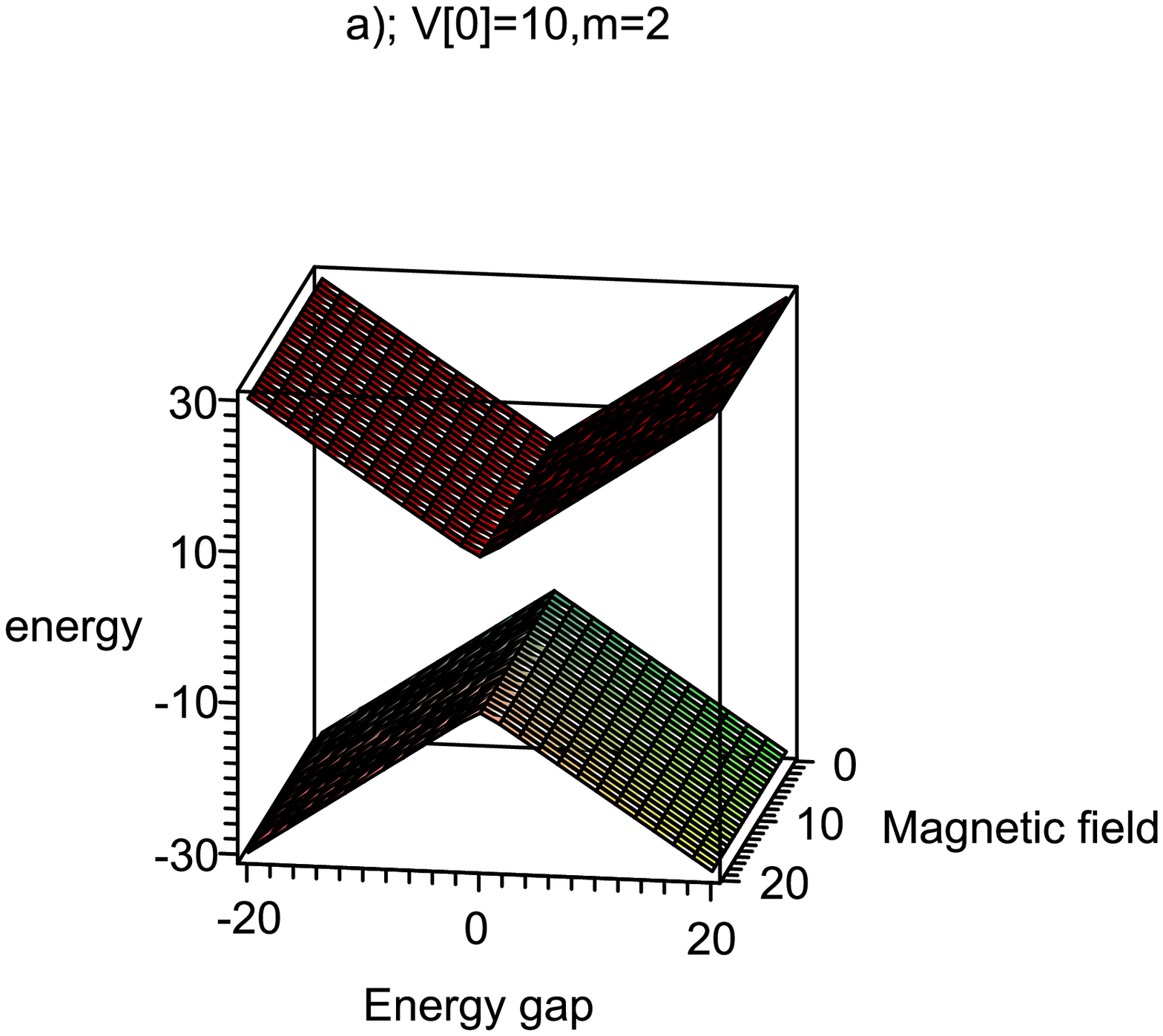}
 \includegraphics[width=2.4in]{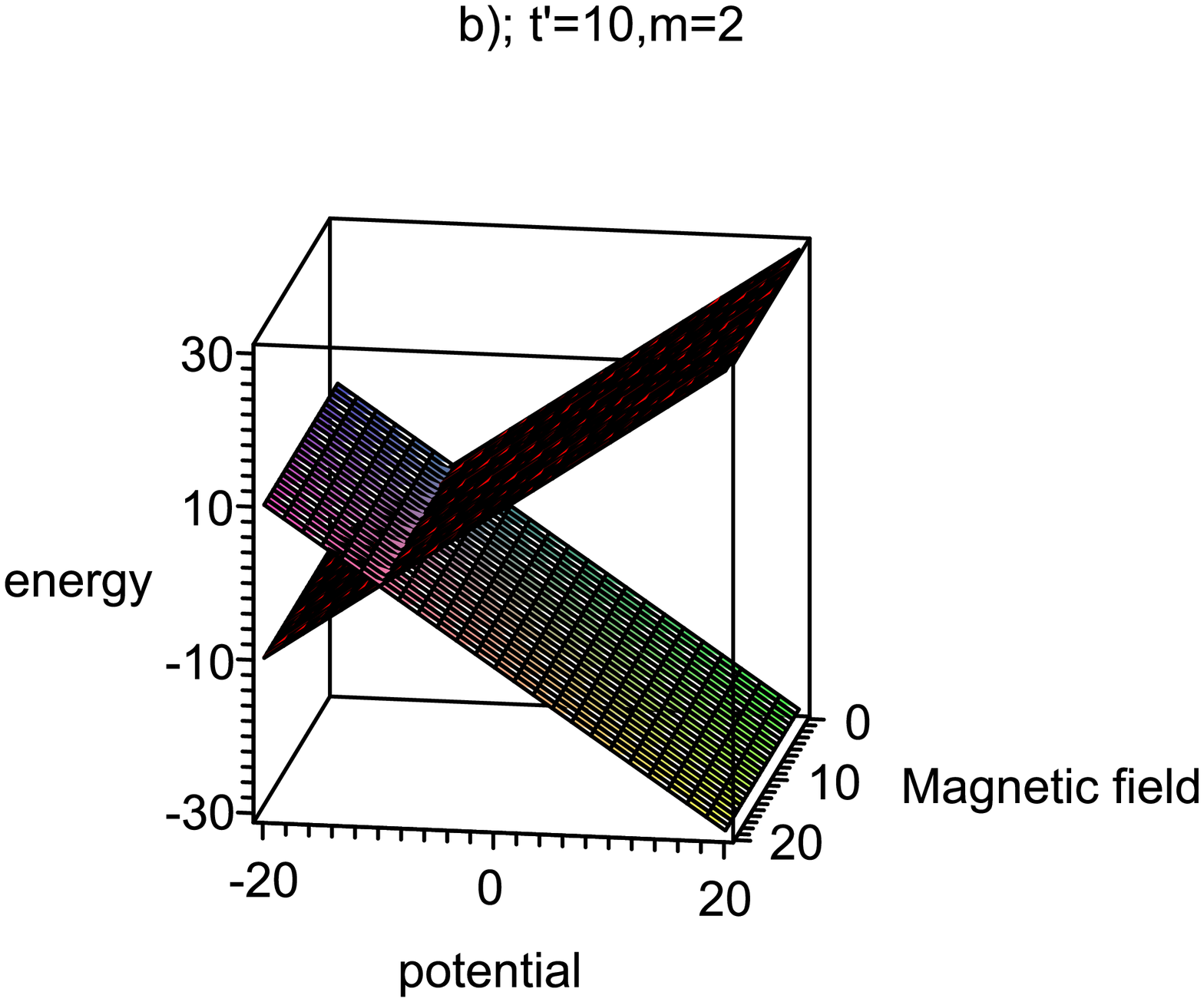}
 \end{center}
 {\sf Figure 12}: %{ \textcolor[rgb]{1.00,0.00,0.00}
 {Energy $E_{m}$ for a
magnetic barrier of width $2d$: a) $E_{m}$ as function of $B'$ and $t'$,  b) $E_{m}$ as function of $B'$ and $V_{0}$.}\\
%\end{figure}
%\end{center}
%%%%%%%%%%%%%%%%%%%%%%%%%%%%%%%%%%%%%%%%%%%%%%%%%%%%%%%%%
\noindent
We notice that in the presence of energy gap, there is a gap separating
the conduction and valence bands, which is missing in the case where $t'=0$ (Figure 12b).
%Dans la présence d'une énergie de gap (t') on vois clairement qu'il y a apparition d'une gap qui sépare la bande de conduction BC et la bande de valence BV %(Figur 12 a) par contre dans l'absence de cet énergie la BC et la BV se touchent (Figure12 b)

To complete the derivation of eigenspinors, we determine the first component $\varphi_{1}'$.
Then, from (\ref{16}) and (\ref{17}) we obtain
\begin{eqnarray}
\varphi_{1}'(x,y)=\frac{-i\hbar\omega_{c}'\sqrt{|m|}}{E_{\sf II}-V_{0}-t'}D_{|m|-1}(x+x_{02})e^{iq_{y}y}.
\end{eqnarray}
After normalization the eigenspinors read as
\begin{eqnarray}\label{ai}
\phi_{{\sf II},m,q_{y}}(x,y)=\frac{1}{\sqrt{2}}\left(%
\begin{array}{c}
  -a_{m}i D_{|m|-1}(x+x_{02}) \\
 b_{m}D_{|m|}(x+x_{02}) \\
\end{array}\
\right)e^{iq_{y}y}
\end{eqnarray}
where the constants are given by
\begin{eqnarray}
a_{m}=s'\sqrt{\frac{E_{\sf II}-V_{0}+s't'}{E_{\sf II}}}, \qquad
b_{m}=\sqrt{\frac{E_{\sf II}-V_{0}-s't'}{E_{\sf II}}}.
\end{eqnarray}
%Equations (\ref{6}) and (\ref{ai}) are the generalization of~\cite{cho}.

As concerning regions {\sf  I} and {\sf III},
the corresponding eigenspinors $\phi_{{\sf I},n,k_{y}}(x,y,x_{01})$ and
$\phi_{{\sf III},n,k_{y}}(x,y,x_{03})$ can be written in compact form
as $\phi_{{\sf I},n,k_{y}}(x,y,x_{0})$ where
\begin{equation}
x_{0}=k_{y}l_{B}^{2}-\left(1-\frac{|B'|}{B}\right)x
\end{equation}
 such that
 $x=-d$ and $x=d$ give $x_{0}=x_{01}$ and
 $x_{0}=x_{03}$, respectively.
%Then the region {\sf I} is
%equivalent to the region {\sf III}.

On the other hand, the energy conservation
between regions {\sf I} and {\sf II} gives
\begin{eqnarray}
E_{\sf I}=E_{\sf II}=E.
\end{eqnarray}
 After replacing, we show that the quantum numbers $n$ and
$m$  verify the relation
\begin{eqnarray}\label{2econ}
\frac{|n|}{|m|}=\frac{|B'|}{B}\frac{E^{2}}{\left(E-V_{0}\right)^{2}-t'^{2}}.
\end{eqnarray}
We have some remarks in order.
In region {\sf I}$\equiv${\sf III} we have $V_{0}=0$ and $t'=0$,
thus (\ref{2econ}) reduces to
\begin{eqnarray}
\frac{|n|}{|m|}=\frac{|B'|}{B}.
\end{eqnarray}
However, in region {\sf II} there two cases:
\begin{eqnarray}
\begin{array}{c}
(E-V_{0})^{2}>t'^{2} \ \ \ \Longrightarrow \ \ \ |m|=+m\\
(E-V_{0})^{2}<t'^{2} \ \ \ \Longrightarrow \ \ \ |m|=-m.
\end{array}
\end{eqnarray}
Finally, the analogue of (\ref{1encon}) is now
given by
\begin{eqnarray}
\frac{E}{V_{0}}=\frac{\sqrt{|n|}}{\sqrt{|n|}-\frac{s'}{s}\sqrt{\frac{|B'|}{B}|m|+\frac{t'^{2}}{\hbar^{2}\omega_{c}^{2}}}}.
\end{eqnarray}
\subsection{Reflexion and transmission coefficients in the presence of $t'$}
%%%%%%%%%%%%%%%%%%%%%%%%%%%%%%%%%%%%%%%%%%%%%%%%%%%%%%%%%%%%%%%%%%%%%%%%%%%%%%%%%

We will see how the results obtained before
can be generalized to the present case. To proceed,
we consider two (barrier) and three regions (diode). % we distinguish between two systems. Indeed,
%as far as two regions are concerned, we
For barrier, we
show that the reflexion and transmission
coefficients are
%Using the same technique for the previous cases in
%magnetic field without $t'$ and distinguishing between
%propagation with positive and negative incidences, one can obtain
%the reflexion and transmission coefficients
\begin{eqnarray}
\rho'(x_{0},x_{02})&=&\frac{\left[sb_{m}A_{\sf I}-a_{m}B_{\sf I}\right]^{2}}
{\left[sb_{m}A_{\sf I}+a_{m}B_{\sf I}\right]^{2}}.\label{23}\\
\tau'(x_{0},x_{02})&=&\frac{4sb_{m}a_{m}C_{\sf I}F_{\sf I}}
{\left[sb_{m}A_{\sf I}+a_{m}B_{\sf I}\right]^{2}}\nonumber\\
&=&\frac{4sb_{m}a_{m}A_{\sf I}B_{\sf I}}
{\left[sb_{m}A_{\sf I}+a_{m}B_{\sf I}\right]^{2}}\label{22}
\end{eqnarray}
%Recall that  $C_{\sf I}F_{\sf I}=A_{\sf I}B_{\sf I}$.
Clearly, what makes
difference with  respect to the former results is the appearance of the
constant parameters $a_m$ and $b_m$.
These coefficients can be used
%Using (\ref{22}) and (\ref{23}), we show that they
to verify the probability
condition
\begin{eqnarray}
\rho'(x_{0},x_{02})+\tau'(x_{0},x_{02})=1.
\end{eqnarray}
%This is the interesting relation, which is analogue to what
%obtained in zero field case, and without $t'$.

%%%%%%%%%%%%%%%%%%%%%%%%%%%%%%%%%%%%%%%%%%%%%%%%%%%%%%%%%%%%%%%%%%%%%%
\begin{center}
  \includegraphics[width=6in]{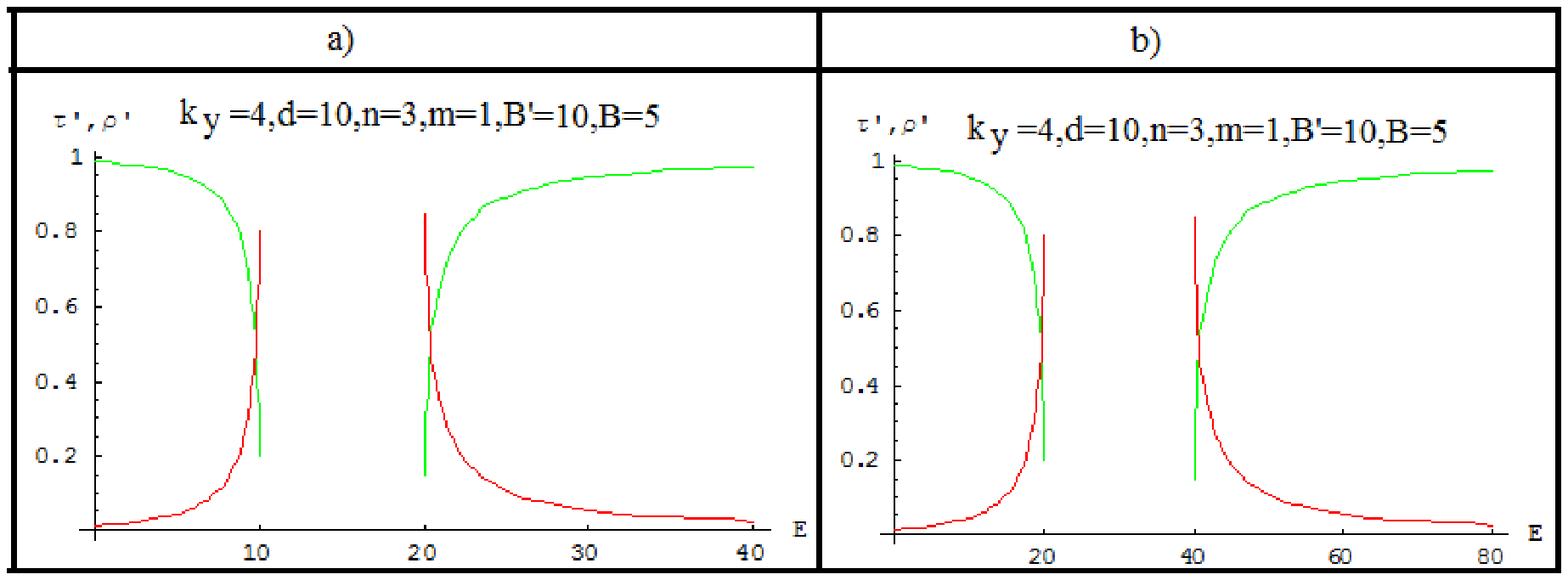}\\
  \end{center}
{\sf Figure 13:} {{ Reflexion $\rho'(x_{0},x_{02})$ (red line) and
transmission $\tau'(x_{0},x_{02})$ (green line) coefficients for a
magnetic barrier of width $2d$ at various energy $E$ for two
cases: a) $V_{0}=15$, $t'=5$ and b) $V_{0}=30$, $t'=10$}}.\\
%%%%%%%%%%%%%%%%%%%%%%%%%%%%%%%%%%%%%%%%%%%%%%%%%%%%%%%%%%%%%%%%%%%%

\noindent {At ($V_{0}=15$, $t'=5$) and
($V_{0}=30$, $t'=10$) the transmission and reflexion profile
(Figure 13) between regions ${\sf I-II}$ and ${\sf II-III}$ show}
{
\begin{itemize}
    \item $-t'<E<t'$: $\tau'(x_{0},x_{02})$ $\rightarrow 1$ and
    $\rho'(x_{0},x_{02})$ $\rightarrow 0$
    \item $t'<E<V_{0}-t'$:  $\tau'(x_{0},x_{02})$ decreases and
    $\rho'(x_{0},x_{02})$ increases.
    \item $V_{0}-t'<E<V_{0}+t'$: there is no transmission and non
    reflexion (not allowed states).
    \item $E>V_{0}+t'$: $\tau'(x_{0},x_{02})$ $\rightarrow 1$ and
    $\rho'(x_{0},x_{02})$ $\rightarrow 0$.
\end{itemize}

As far as three regions are concerned, one can inspire
from the case without gap to obtain the reflection and
transmission amplitudes, such as
%obtain
\begin{eqnarray}
\textbf{\textcolor[rgb]{1.00,0.00,0.00}{R'}}&=&\frac{\rho'(d_{1},d_{2})+\rho'(d_{3},d_{2}')
-2r^{+}_{n,m}(d_{1},d_{2})r^{+}_{n,m}(d_{3},d_{2}')}
{1+\rho'(d_{1},d_{2})\rho'(d_{3},d_{2}')
-2r^{+}_{n,m}(d_{1},d_{2})r^{+}_{n,m}(d_{3},d_{2}')} \\
%\end{eqnarray}
%\begin{eqnarray}
\textbf{\textcolor[rgb]{1.00,0.00,0.00}{T'}} &=&\frac{1+\rho'(d_{1},d_{2})\rho'(d_{2}',d_{3})
-\rho'(d_{1},d_{2})-\rho'(d_{2}',d_{3})}
{1+\rho'(d_{1},d_{2})\rho'(d_{2}',d_{3})
-2r^{+}_{n,m}(d_{1},d_{2})r^{+}_{n,m}(d_{2}',d_{3})}.
\end{eqnarray}
After a straightforward calculation, we find
\begin{eqnarray}
\textbf{\textcolor[rgb]{1.00,0.00,0.00}{R'+ T'=1}}.
\end{eqnarray}
%This also has been obtained by studying the Dirac particles
%through an strip of graphene, the length $d$ and the width $W$,
%confine bey an inhomogeneous magnetic fields. For
Note that, for $B=B'$ we discover the results obtained
in~\cite{cho}, which  shows that our results are more generals.

%%%%%%%%%%%%%%%%%%%%%%%%%%%%%%%%%%%%%%%%%%%%%%%%%%%%%%%
\section{Klein Paradox} %for particles in graphene }
%%%%%%%%%%%%%%%%%%%%%%%%%%%%%%%%%%%%%%%%%%%%%%%%%%%%%%%

We complete the present work by analyzing the Klein paradox
for the present system. This can be done by introduce other considerations based on the current-carrying states
 and study different limiting cases.

%%%%%%%%%%%%%%%%%%%%%%%%%%%%%%%%%%%%%%%%%%%%%%%%%%%%%%%%%%%%%%%%%%%%
\subsection{Propagation from region {\sf I} to region {\sf II}: $(x=-d)$}
%%%%%%%%%%%%%%%%%%%%%%%%%%%%%%%%%%%%%%%%%%%%%%%%%%%%%%%%%%%%%%%%%%%%

We consider the scattering of a Dirac fermion of energy $E$ from
an electrostatic step-function potential as shown in Figure 14.
This problem is an archetype problem in nonrelativistic quantum
mechanics. For relativistic quantum mechanics, we will find that
the solution leads to a paradox (Klein paradox) when the potential
is strong.

%%%%%%%%%%%%%%%%%%%%%%%%%%FIGURE%%%%%%%%%%%%%%%%%%%%%%%%%%%%%%%%%%%%
\begin{center}
  \includegraphics[width=2.2in]{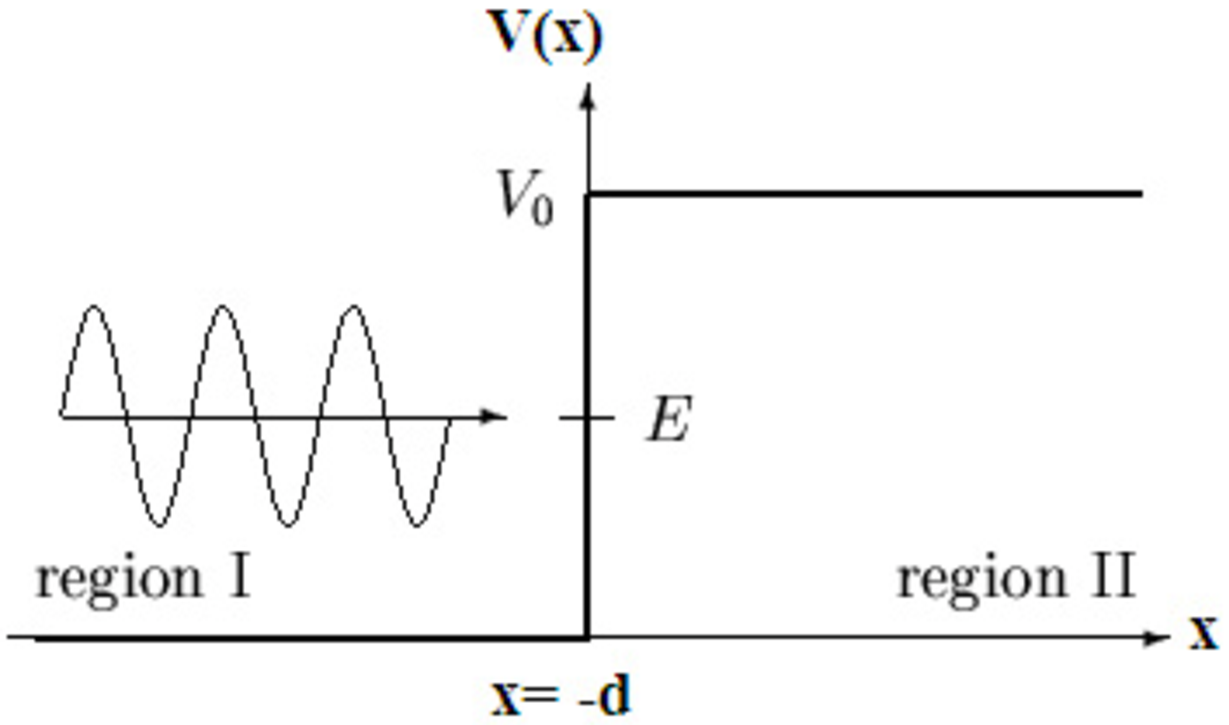}
  \end{center}
{\sf Figure 14:} {Electrostatic potential idealized with a sharp
boundary, with an incident free scalar fermion of energy moving
to the right in region {\sf I}.}\\
%%%%%%%%%%%%%%%%%%%%%%%%%%%%%%%%%%%%%%%%%%%%%%%%%%%%%%%%%%%%%%%%%%%%%%%%%%%%%%%%%%%%%%%%

According to the previous analysis for two regions,
it is easy to note that the solution of the Dirac equation in the region
{\sf I} and {\sf II} are given by
\begin{eqnarray}
&& \phi_{x<-d}=\frac{1}{\sqrt{2}}\left(%
\begin{array}{c}
  -si D_{|n|-1}(x+x_{01}) \\
 D_{|n|}(x+x_{01}) \\
\end{array}%
\right)+\frac{1}{\sqrt{2}}R\left(%
\begin{array}{c}
  -si D_{|n|-1}(-x-x_{01}) \\
 D_{|n|}(-x-x_{01}) \\
\end{array}%
\right)
\\
&& \phi_{x>-d}=\frac{1}{\sqrt{2}}T\left(%
\begin{array}{c}
  -a_{m}i D_{|m|-1}(x+x_{02}) \\
 b_{m}D_{|m|}(x+x_{02}) \\
\end{array}%
\right)
\end{eqnarray}
where $R$ and $T$ are reflected and transmitted coefficients,
respectively.
Imposing the boundary condition that $\phi$ be continuous at
$(x=-d)$ gives the relation
\begin{eqnarray}
\left(%
\begin{array}{c}
  s D_{|n|-1}(-d+x_{01}) \\
 D_{|n|}(-d+x_{01}) \\
\end{array}%
\right)+R\left(%
\begin{array}{c}
  s D_{|n|-1}(d-x_{01}) \\
 D_{|n|}(d-x_{01}) \\
\end{array}%
\right)=T\left(%
\begin{array}{c}
  a_{m}D_{|m|-1}(-d+x_{02}) \\
 b_{m}D_{|m|}(-d+x_{02}) \\
\end{array}%
\right).
\end{eqnarray}
Solving for $R$ and $T$ to obtain
\begin{eqnarray}
R&=&\frac{sb_{m}u_{1}v_{3}-a_{m}v_{1}u_{3}}{a_{m}v_{2}u_{3}-sb_{m}u_{2}v_{3}}\\
T&=&\frac{s(u_{1}v_{2}-v_{1}u_{2})}{a_{m}v_{2}u_{3}-sb_{m}u_{2}v_{3}}
\end{eqnarray}
where we use the notation
\begin{eqnarray}
&&u_{1}=D_{|n|-1}(-d+x_{01}), \qquad u_{2}=D_{|n|-1}(d-x_{01}),
\qquad u_{3}=D_{|m|-1}(-d+x_{02}) \nonumber\\
&&v_{1}=D_{|n|}(-d+x_{01}),
\qquad v_{2}=D_{|n|}(d-x_{01}), \qquad v_{3}=D_{|m|}(-d+x_{02}).
\end{eqnarray}

To proceed further, we introduce the current-carrying states. This is based on the
current associated to Dirac equation, which is
\begin{eqnarray}
J=ev\sum_{i}\phi^{+}\sigma_{i}\phi
\end{eqnarray}
where $i=x,y$. As an immediate application, the incident current is given by
\begin{eqnarray}
J_{\sf I}=ev\left(\phi^{+}_{\sf I}\sigma_{x}\phi_{\sf I}+\phi^{+}_{\sf I}
\sigma_{y}\phi_{\sf I}\right)=sevu_{1}v_{1}.
\end{eqnarray}
This can be used to evaluate
the final currents to the left and right of the potential boundary, which read as
\begin{eqnarray}
J_{x<-d}& = &sev\left(u_{1}+Ru_{2}\right)\left(v_{1}+Rv_{2}\right)\\
J_{x>-d} &=& \frac{ev}{2}|T|^{2}u_{3}v_{3}\left[i(a_{m}^{+}b_{m}-a_{m}b_{m}^{+})+(a_{m}^{+}b_{m}+a_{m}b_{m}^{+})\right]
\label{posij}.
\end{eqnarray}
(\ref{posij}) can be split  into three
parts
\begin{eqnarray}\label{spljp}
J_{x>-d}  = \left\{%
\begin{array}{ll}
    -ev|T|^{2}a_{m}b_{m}u_{3}v_{3}, &\qquad \hbox{$V_{0}>E+t'$} \\
    -iev|T|^{2}a_{m}b_{m}u_{3}v_{3}, &\qquad \hbox{$E-t'<V_{0}<E+t'$} \\
    ev|T|^{2}a_{m}b_{m}u_{3}v_{3}, &\qquad \hbox{$V_{0}<E-t'$} \\
\end{array}%
\right.
\end{eqnarray}
with the condition $a_{m}b_{m}>0$.

Recall that, the reflexion and transmission
amplitudes are related to the currents through
%Now using the relation between different coefficients
%and the current to obtain
%The transmission coefficient is
\begin{equation}
\mathbf{R}=\frac{J_{\sf I}-J_{x<-d}}{J_{\sf I}}, \qquad
\mathbf{T}=\frac{J_{x>-d}}{J_{\sf I}}.
\end{equation}
After replacing, we end up with
\begin{eqnarray}
\mathbf{R}&=&\frac{u_{1}v_{1}\left(a_{m}v_{2}u_{3}-sb_{m}u_{2}v_{3}\right)^{2}-a_{m}b_{m}u_{3}v_{3}\left(u_{1}v_{2}
-v_{1}u_{2}\right)^{2}}{u_{1}v_{1}\left(a_{m}v_{2}u_{3}-sb_{m}u_{2}v_{3}\right)^{2}}\\
\mathbf{T}&=&\frac{u_{3}v_{3}\left[i(a_{m}^{+}b_{m}-a_{m}b_{m}^{+})+(a_{m}^{+}b_{m}+a_{m}b_{m}^{+})\right]\left(u_{1}v_{2}
-v_{1}u_{2}\right)^{2}}{2u_{1}v_{1}\left(a_{m}v_{2}u_{3}-sb_{m}u_{2}v_{3}\right)^{2}}.
\end{eqnarray}
These show that
the probability is
\begin{equation}\label{lapbb}
\mathbf{R}+ \mathbf{T}=1.
\end{equation}
We can inspect (\ref{spljp}) further to derive other results.
This can be achieved by considering three interesting
cases.

%%%%%%%%%%%%%%%%%%%%%%%%%%%%%%%%%%%%%%%%%%%%%%%%
\subsection{Limiting cases}
%%%%%%%%%%%%%%%%%%%%%%%%%%%%%%%%%%%%%%%%%%%%%

In region {\sf II} there are three distinct cases, depending on
the strength of the potential. This is shown by Figure 15:

%%%%%%%%%%%%%%%%%%%%%%%%%%%%%%%%FIGURE%%%%%%%%%%%%%%%%%%%%%%%
\begin{center}
  \includegraphics[width=2.2in]{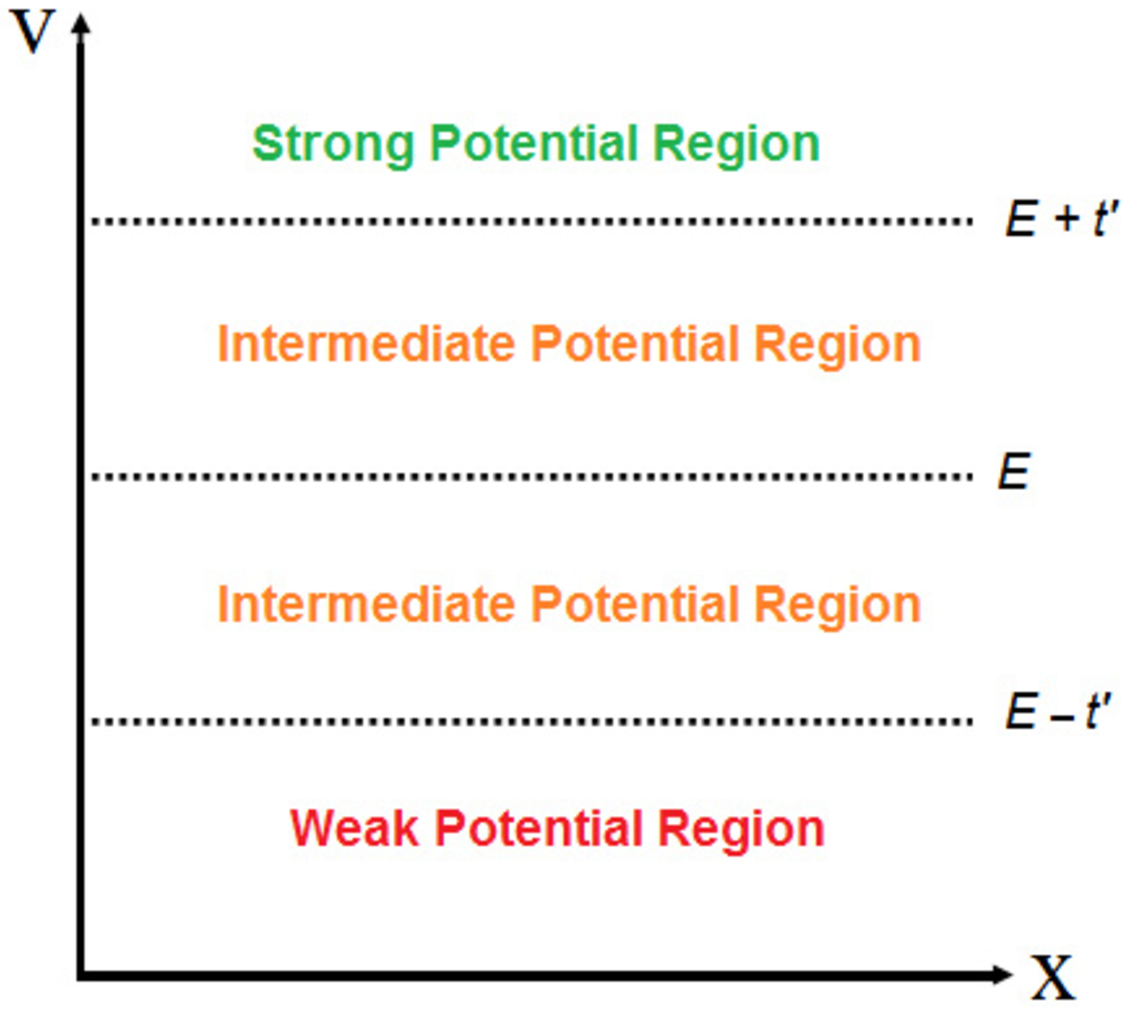}\\
  {\sf Figure 15:} Energy level diagram for a fermion in region
{\sf II}.
\end{center}

%%%%%%%%%%%%%%%%%%%%%%%%%%%%%%%%%%%%%%%%%%%%%%%%%%%%%%%%%%%%%%%%%%

%We see that $|m|$ can be $+m$ or $|-m|$ depending on the strength
%of the potential $V_{0}$.

Let us analyze each case separately and underline its
physical properties. Indeed, in the {weak potential} that
corresponds to %$V_{0}<E-t' \Longrightarrow
$E-V_{0}>t'$, we have
a restriction on the quantum numbers and parameter constants, such as
%, we have the configuration
 $$ |m|=+m,
 \qquad  a_{m}^{\dagger}=a_{m}, \qquad b_{m}^{\dagger}=b_{m}.$$
In such case, the  reflexion and
transmission  amplitudes are given by
\begin{eqnarray}
&& \mathbf{R}=\frac{u_{1}v_{1}\left(a_{m}v_{2}u_{3}-sb_{m}u_{2}v_{3}\right)^{2}-a_{m}b_{m}u_{3}v_{3}\left(u_{1}v_{2}
-v_{1}u_{2}\right)^{2}}{u_{1}v_{1}\left(a_{m}v_{2}u_{3}-sb_{m}u_{2}v_{3}\right)^{2}}\\
&& \mathbf{T}=\frac{a_{m}b_{m}u_{3}v_{3}\left(u_{1}v_{2}
-v_{1}u_{2}\right)^{2}}{u_{1}v_{1}\left(a_{m}v_{2}u_{3}-sb_{m}u_{2}v_{3}\right)^{2}}.
\end{eqnarray}
%\begin{eqnarray}
%\mathbf{T}+\mathbf{R}=1
%\end{eqnarray}
They verify the probability condition (\ref{lapbb}).
Thus the incident beam is partly reflected and partly transmitted.
This is similar to the result obtained in nonrelativistic quantum
mechanics. The last expression shows that the total probability is
conserved.

As far as
the {intermediate potential} is concerned, i.e.
%$E-t'<V_{0}<E+t' \Longrightarrow $
$|E-V_{0}|<t'$, different quantities
reduce as
%characterized by
 $$ |m|=-m, \qquad a_{m}^{\dagger}=-a_{m},\qquad  b_{m}^{\dagger}=b_{m}.$$
%For the case of the \textbf{intermediate potential} $(|m|=-m)$,
The corresponding amplitudes are
\begin{eqnarray}
\mathbf{R}&=&\frac{u_{1}v_{1}\left(a_{m}^{2}v_{2}^{2}u_{3}^{2}-b_{m}^{2}u_{2}^{2}v_{3}^{2}\right)^{2}+ia_{m}b_{m}u_{3}v_{3}\left(u_{1}v_{2}
-v_{1}u_{2}\right)^{2}}{u_{1}v_{1}\left(a_{m}^{2}v_{2}^{2}u_{3}^{2}-b_{m}^{2}u_{2}^{2}v_{3}^{2}\right)^{2}}\\
\mathbf{T}&=&-\frac{ia_{m}b_{m}u_{3}v_{3}\left(u_{1}v_{2}
-v_{1}u_{2}\right)^{2}}{u_{1}v_{1}\left(a_{m}^{2}v_{2}^{2}u_{3}^{2}+b_{m}^{2}u_{2}^{2}v_{3}^{2}\right)}
\end{eqnarray}
%\begin{eqnarray}
%\mathbf{T}+\mathbf{R}=1
%\end{eqnarray}
where
the probabilities  sum to unity, as
must be the case, since reflection and transmission are the only
possible outcomes for a fermion incident on the barrier.

 In the {strong potential} case, i.e.
 %V_{0}>E+t' \Longrightarrow
 $|E-V_{0}|>t'$, we have
 $$|m|=-m,\qquad
 a_{m}^{\dagger}=-a_{m},\qquad  b_{m}^{\dagger}=-b_{m}$$
%For the case of the \textbf{strong potential} $(|m|=-m)$, the
%transmission and reflection coefficients are
which is showing
\begin{eqnarray}
&& \mathbf{R} = \frac{u_{1}v_{1}\left(a_{m}b_{m}u_{3}v_{3}\left(u_{1}v_{2}
-v_{1}u_{2}\right)^{2}+a_{m}v_{2}u_{3}-sb_{m}u_{2}v_{3}\right)^{2}}{u_{1}v_{1}\left(a_{m}v_{2}u_{3}+b_{m}u_{2}v_{3}\right)^{2}}>1\\
&& \mathbf{T} = -\frac{a_{m}b_{m}u_{3}v_{3}\left(u_{1}v_{2}
-v_{1}u_{2}\right)^{2}}{u_{1}v_{1}\left(a_{m}v_{2}u_{3}+b_{m}u_{2}v_{3}\right)^{2}}<0.\label{bigT}
\end{eqnarray}
%\begin{eqnarray}
%\mathbf{T}+\mathbf{R}=1
%\end{eqnarray}
The probability is still conserved, but only at the cost of a
negative transmission amplitude and a reflection amplitude
which exceeds unity. The strong potential appears to give rise to
a paradox. There is no paradox if we consider that in the strong
potential case the potential is strong enough to create
particle-antiparticle pairs. The antiparticles are attracted by
the potential and create a negative charged current moving to the
right. This is the origin of the negative transmission
amplitude, i.e. (\ref{bigT}). The particles, on the other hand, are reflected from
the barrier and combined with the incident particle beam (which is
completely reflected) leading to a positively charged current,
moving to the left and with magnitude greater than that of the
incident beam.

%%%%%%%%%%%%%%%%%%%%%%%%%%%%%%%%%%%%%%%%%%%%%%%%%%%%%%%
\section{Conclusion}
%%%%%%%%%%%%%%%%%%%%%%%%%%%%%%%%%%%%%%%%%%%%%%%%%%%%%%%

We considered a system composed of different regions of
Dirac fermions in the presence of an inhomogeneous magnetic field %$(B,B')$
and confining potential $V(x)$ in one-direction.
The energy spectrum solutions are obtained in terms of different
parameters and quantum numbers for each regions. To underline
some physical properties of the obtained solutions, we analyzed the
energy conservation. This allowed us to establish
interesting relations and therefore solve some issues
related to reflexion and transmission of the system.

More precisely, by considering our system as
as barrier, we derived interesting results. In fact,
using the continuity equation at different points we explicitly
determined the reflexion and transmission coefficients. These
are used to define the  corresponding amplitudes and therefore to
show that their probabilities sum to unity. Different cases
are treated, which concerned total reflecting and transmitting beams
where they are interpreted as mirror or diopter systems.

Subsequently, we focussed on three regions of two fixed points
$d$ and $-d$. Writing the continuity at each point, we
derived different quantities those
are needed to characterize the beam of the present system.
Indeed, we reached the conclusions that
the probabilities of reflection and transmission
%In other words, the probabilities of reflection and transmission
sum to unity, as must be the case, since they are the only possible outcomes for a fermion
incident on the barrier.
%\item This is among the interesting conclusion
%reached in this section.
%Furthermore,
Furthermore, (\ref{82}) and (\ref{83}) yielded that under resonance
conditions: $|n|=|m|$ and $s=s'$,
 the barrier becomes transparent, i.e. $T=1$. More significantly, however, the barrier remains always
perfectly transparent for $|n|=|m|$. This latter is the feature
unique to massless Dirac fermions
and directly related to the
Klein paradox in quantum electrodynamics.

Another interesting cases is analyzed, which concerned introducing a gap.
After getting the energy spectrum solutions, we discussed different issues
and among them the energy conservation. This allowed us to generalize the
former analysis to gap case. As interesting results, we showed that
the probabilities of reflecting and transmitting amplitudes sun to unity
as well. Requiring that $B=B'$, we recovered the result obtained in~\cite{cho}.

Finally, we discussed the Klein paradox by involving the
 current-carrying states for different regions. Using their relations to the
 reflexion and transmission amplitudes, we %These allowed us
%firstly to
checked the probability by evaluating different quantities. Moreover,
we treated three different limiting cases, which
concern  week, intermediate and strong potentials. For two last cases,
the transmission amplitude is obtained with a negative sign, however
when it is added to the reflexion coefficient gives a sum equal unit.

%%%%%%%%%%%%%%%%%%%%%%%%%%%%%%%%%%%%%%%%%%%%%%%%%%%%%%%
\section*{Acknowledgment}
%%%%%%%%%%%%%%%%%%%%%%%%%%%%%%%%%%%%%%%%%%%%%%%%%%%%%%%

The authors are thankful to Dr. El Bouazzaoui Choubabi for fruitful
discussions about the tunneling effect in graphene.

\end{document}